\documentclass[footinbib,twocolumn,aps,prl,amsmath,superscriptaddress,notitlepage,longbibliography]{revtex4-1}
\usepackage{amssymb}
\usepackage{mathrsfs}
\usepackage{graphicx}
\usepackage{float}
\usepackage[caption=false]{subfig}
\usepackage{epstopdf}

\usepackage[normalem]{ulem}
\usepackage{verbatim}
\usepackage{xcolor}
\usepackage{bm}

\usepackage[utf8x]{inputenc}

\def\blue{\textcolor{blue}}
\def\red{\textcolor{red}}

\begin{document}

\def\qv{\vec{q}}
\def\red{\textcolor{red}}
\def\blue{\textcolor{blue}}
\def\magenta{\textcolor{magenta}}
\def\apricot{\textcolor{Apricot}}

\def\GJ{\textcolor{black}}
\def\TT{\textcolor{ForestGreen}}
\definecolor{ora}{rgb}{1,0.45,0.2}
\def\LH{\textcolor{black}}

\newcommand{\norm}[1]{\left\lVert#1\right\rVert}
\newcommand{\ad}[1]{\text{ad}_{S_{#1}(t)}}

\title{Many-Body Topological and Skin States without Open Boundaries}

\author{Ching Hua Lee}
\email{phylch@nus.edu.sg}
\affiliation{Department of Physics, National University of Singapore, Singapore 117542}

\date{\today}

\date{\today}
\begin{abstract}
Robust boundary states have been the focus of much recent research, both as topologically protected states and as non-Hermitian skin states. In this work, we show that many-body effects can also induce analogs of these robust states in place of actual physical boundaries. Particle statistics or suitably engineered interactions i.e. in ultracold atomic lattices can restrict the accessible many-body Hilbert space, and introduce effective boundaries in a spatially periodic higher-dimensional configuration space. We demonstrate the emergence of topological chiral modes in a two-fermion hopping model without open boundaries, with fermion pairs confined and asymmetrically propagated by suitably chosen fluxes. Heterogeneous non-reciprocal hoppings across different particle species can also result in robust particle clumping in a translation invariant setting, reminiscent of skin mode accumulation at an open boundary. But unlike fixed open boundaries, effective boundaries correspond to the locations of impenetrable particles and are dynamic, giving rise to fundamentally different many-body vs. single-body time evolution behavior. Since non-reciprocal accumulation is agnostic to the dimensionality of restricted Hilbert spaces, our many-body skin states generalize directly in the thermodynamic limit. The many-body topological states, however, are nontrivially dimension-dependent, and their detailed exploration will stimulate further studies in higher dimensional topological invariants.
\end{abstract}

\maketitle

\noindent{\textit{Introduction.--}} Much of contemporary condensed matter research have revolved round robust boundary phenomena. Topological boundary states are anomaly manifestations of nontrivial bulk topology, and have have been extensively investigated in quantum spin and anomalous Hall (Chern) insulators~\cite{kane2005z,bernevig2006quantum,konig2007quantum,liu2008quantum,roy2009topological,qi2011topological,chang2013experimental,he2017chiral,tang2019three}, nodal semimetals~\cite{lv2015experimental,soluyanov2015type,ezawa2016loop,bzduvsek2016nodal,bian2016topological,neupane2016observation,zhong2017three,zhang2019quantum,lee2019enhanced,zhang2019cyclotron} and various topological metamaterials~\cite{lu2013weyl,peano2015topological,yang2015topological,nash2015topological,susstrunk2015observation,fleury2016floquet,meeussen2016geared,lin2017line,zhang2017topological,khanikaev2017two,lee2018topological,mao2018maxwell,ma2019topological,ozawa2019topological}, some with potential applications in electronics and photonics~\cite{mellnik2014spin,zhang2015electrical,zhang2016topologicalIC,bandres2018topological,longhi2018non,wang2019topologically,zeng2020electrically}. More recently, non-Hermitian skin effect (NHSE) boundary states which accumulate from unbalanced gain/loss have also seen much experimental~\cite{tobias2019observation,hofmann2019reciprocal,xiao2020non,ananya2019observation,weidemann2020topological} and theoretical~\cite{Lee2016nonH,Yao2018nonH2D,lee2018tidal,song2019non,song2019realspace,jiang2019interplay,kunst2019non,Lee2019hybrid,li2019geometric,borgnia2020nonH,zhang2019correspondence,li2019topology,yoshida2019mirror,yang2019auxiliary,longhi2019probing,luo2020skin,yi2020non,PhysRevLett.124.086801,PhysRevLett.124.066602} advances, fueled by various intriguing implications like modified bulk-boundary correspondences~\cite{xiong2018does,kunst2018biorthogonal,yao2018edge,yokomizo2019non,lee2019anatomy}, critical behavior~\cite{mu2019emergent,li2020critical,liu2020helical}, discontinuous band geometry~\cite{lee2019unraveling} and unconventional responses~\cite{lee2019unraveling,lee2020ultrafast,schomerus2020nonreciprocal}, alongside possible sensing applications~\cite{schomerus2020nonreciprocal,mcdonald2020exponentially,budich2020non,weidemann2020topological}.

While both topological and skin modes mathematically arise from the breaking of translation invariance, this translation need not be in physical real space. In particular, many-body effects like Pauli exclusion and interactions can restrict the accessible Hilbert space and break the ``translation'' invariance of the many-body configuration space. A simplest illustration involves two (distinguishable) fermions on a line, where Pauli exclusion fixes their relative ordering and partitions the configuration space into two disjoint halves.  More generally, diverse types of effective boundaries and localized inhomogeneities can be engineered through suitable interactions. Indeed, interesting new physics have been known to emerge when interactions constrain the accessible Hilbert space, as epitomized by Fractional quantum Hall states with non-abelian quasi-particles determined by their detailed pseudopotential profiles~\cite{haldane1985finite,simon2007pseudopotentials,ardonne2009domain,barkeshli2009structure,davenport2012multiparticle,lee2015geometric,yang2017generalized,lee2018floquet,yang2019emergent}. 

In this work, we investigate how skin and topological states, conventionally regarded as boundary phenomena, can emerge from many-body interactions in a purely periodic boundary condition (PBC) lattice setting. Unlike static physical boundaries, their effective ``boundaries'' are marked by degenerate configurations, and themselves depend on the particle positions. Their dramatically different dynamics and spectra manifest as emergent hopping non-locality for PBC skin states, and emergent asymmetric propagation for PBC chiral Chern modes.

\noindent\textit{``Boundaries'' via restricting many-body Hilbert spaces.--} The many-body configuration space of a generic system contains subspaces where two or more particles occupy the same state. If particle statistics i.e. Pauli exclusion or specially designed interactions render them inaccessible, these subspaces will serve as effective ``boundaries'' of the configuration space, even in the absence of any physical boundary i.e. edge or surface terminations. To engineer such boundaries, we shall use density-dependent lattice hoppings that are attenuated at high occupancies, such that transitions into certain degenerate states vanish.

\noindent\textit{Interaction-induced skin ``boundary'' states.--} We first present a simple periodic 1D monoatomic lattice with pronounced interaction-induced ``boundary'' effects. Consider an interacting periodic chain with two or more species of bosons $\sigma$:
\begin{eqnarray}
H_\text{1D}&=&\sum_{x,\sigma}t^+_{\sigma}c^\dagger_{x+1,\sigma}c_{x,\sigma}+t^-_{\sigma}c^\dagger_{x,\sigma}c_{x+1,\sigma}\notag\\
&&-\left(V^+_{\sigma}c^\dagger_{x+1,\sigma}c_{x,\sigma}\rho_{x+1}+V^-_{\sigma}c^\dagger_{x,\sigma}c_{x+1,\sigma}\rho_x\right),\qquad
\label{H1D}
\end{eqnarray}
with $c^\dagger_{x,\sigma}$($c_{x,\sigma}$) creating(annihilating) a $\sigma$ boson at site $x$, and $\rho_x=\sum_\sigma c^\dagger_{x,\sigma}c_{x,\sigma}$ the density operator across all species. $H_\text{1D}$ may be approximately simulated by a chain of ultracold fermions or repulsive bosons~\cite{lu2012quantum,aikawa2014reaching,de2019degenerate}. 

To understand the significance of effective boundaries in $H_\text{1D}$ and how they can be induced by suitable interactions, we first review its non-interacting limit of $V^\pm_\sigma=0$, where it reduces to the (multi-component) Hatano-Nelson model~\cite{HN1996prl}. Generically, its species-dependent asymmetric hopping amplitudes $|t_\sigma^+|\neq |t_\sigma^-|$ causes \emph{all} states to evolve by asymmetrically growing in the direction larger hopping~\cite{yao2018edge,yokomizo2019non,lee2019anatomy}. If the system is periodic, any initial state will be amplified indefinitely as it repeatedly circumnavigates the chain, giving rise to complex eigenenergies. But under open boundary conditions (OBCs), all states ultimately accumulate at one end of the chain and cannot be amplified further, resulting in real eigenenergies. Indeed, the \emph{entire} spectrum is drastically modified by the boundary conditions, even in the thermodynamic limit i.e. the NHSE. Its corresponding eigenstates are all exponentially localized near the open boundary with inverse decay lengths/skin depths given by $\log\left|\frac{t^+_\sigma}{t^-_\sigma}\right|$, and are thus known as skin states~\cite{yao2018edge}.

In $H_\text{1D}$, the role of the $V^\pm_\sigma$ interactions is to dynamically destructively interfere with the asymmetric $t^\pm_\sigma$ hoppings whenever the destination site already contains $N_x>0$ particles (of any species). Assuming that $n^\pm_\sigma=t^\pm_\sigma/V^\pm_\sigma \in \mathbb{R}^+$, the effective hoppings onto site $x$ will be modified from $t^\pm_\sigma$ to $t^\pm_\sigma - V^\pm_\sigma N_x=t^\pm_\sigma\left(1-N_x/n^\pm_\sigma\right)$, which are weaker than the single-particle hoppings unless $ N_x>2n^\pm_\sigma$. By varying the relative values of 
$n^\pm_\sigma$ for different hopping directions $\pm$ and species $\sigma$, one can obtain a rich array of competitive or cooperative behaviors where a particular particle distribution can simultaneously promote and inhibit the transfer of the various species.
In particular, if we set $n^\pm_\sigma$ to a fixed integer $n$, the hopping onto a site already containing $n$ bosons vanishes. The accessible Hilbert space is thus restricted to the subspace with at most $n$ bosons per site, provided no site contains $n+1$ or more bosons initially. %
\thickmuskip=3mu
\begin{figure}
\includegraphics[width=1\linewidth]{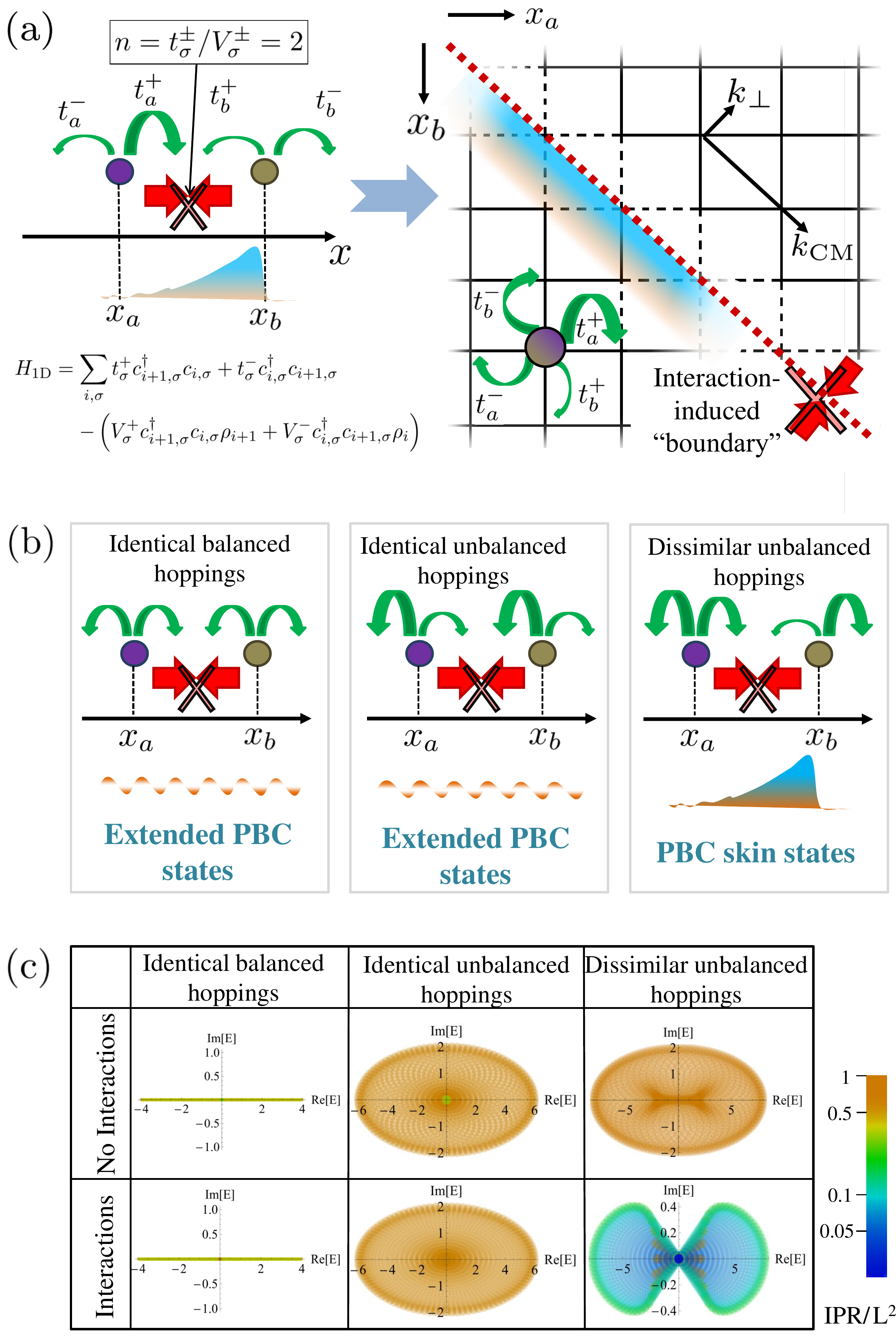}
\caption{(a) (Left) Our interacting chain $H_\text{1D}$ has asymmetric hoppings which vanish if they bring $n+1=2$ particles together. Under PBCs, interaction-induced particle clusters correspond to ``boundary'' skin states along the inaccessible subspace $x_a=x_b$ in the 2-body configuration space. (b) While ordinary OBC skin states require only unbalanced hoppings, PBC skin states also require them to be different across the species. (c) Energy spectra and localization of PBC skin eigenstates $\psi(x)$ in the $L\times L=50^2$-site configuration space, as quantified by their $\text{IPR}=(\sum_{\bold x} |\psi(\bold x)|^4)^{-1}$. $\text{IPR}\sim L^2$ for an extended state (brown), while $\text{IPR}\sim 1$ for full localization (blue). Interactions only induce extensively localized skin states when the hoppings are both unbalanced and dissimilar across species. Hoppings are $t_{1,2}^\pm=1$ (Left), $t^-_{a,b}=2, t^+_{a,b}=1$ (Middle) and $t^-_a=3,t^-_b=1$, $t^+_{a,b}=2$ (Right).  }
\label{fig:skin2}
\end{figure}

\noindent\textit{$N=2$ toy example.--} We next detail the simplest case with $N=2$ bosons of different species on a PBC chain with $L$ sites. Its 2D configuration space is indexed by tuples $(x_a,x_b)\in \{1,...,L\}^2$, which represent the positions of bosons $a$ and $b$ [Fig.~\ref{fig:skin2}a]. Horizontal/vertical hoppings in this space correspond to the interaction-modified effective hoppings of bosons $a$/$b$. Without interactions, the system is a tensor product of two decoupled rings, and every configuration is allowed.  But with interactions set to the simplest value of $n=t^\pm_\sigma/V^\pm_\sigma=1$, particles can no longer hop onto an already occupied site, and hence no site will be doubly occupied~\footnote{If the two particles were initially at the same site, they will no longer be so once they start to evolve.}. This removes the ``diagonal'' subspace $x_a=x_b$ from the accessible Hilbert space[Fig.~\ref{fig:skin2}a]. As illustrated, $x_a=x_b$ behaves like an open ``boundary'' of the configuration space, and changes its topology from a torus to a (45$^\circ$ rotated) cylinder~\footnote{If OBCs were used instead of PBCs, the configuration space would have consisted of two disconnected triangles corresponding to $x_a<x_b$ and $x_a>x_b$.}.

The interaction-induced $x_a=x_b$ configuration space ``boundary'' will host skin states whenever its perpendicular hoppings induce the NHSE, even when the physical lattice satisfies PBCs. To investigate when that can occur, 
we re-express our 2-particle Hamiltonian as a \emph{single-particle} Hatano-Nelson chain in the (1,-1) direction perpendicular to the $x_a=x_b$ line, with effective hoppings depending on the center-of-mass (CM) momentum $k_\text{CM}$:
\begin{eqnarray}
H_\text{1D}^{N=2}(\bold k)&=&\sum_{\pm} t^\pm_a e^{\pm ik_a}+t^\pm_b e^{\pm ik_b}
=\sum_\pm T_\perp^\pm e^{\pm ik_\perp},\qquad
\label{H1D2}
\end{eqnarray}
where~\footnote{Both momentum components $k_\text{CM}$ and $k_\perp$ are well-defined since the system possesses PBCs.} $k_{a,b}=k_\text{CM}\pm k_\perp$ and $T^\pm_\perp=t_a^\pm e^{\pm i k_\text{CM}}+t_b^\mp e^{\mp i k_\text{CM}}$ are the effective right/left hoppings along $k_\perp$. As long as $|T_\perp^+|\neq|T_\perp^-|$, PBC skin states will accumulate at the interaction-induced  $x_a=x_b$ ``boundary''. 
This is equivalent to the following inequality on the $t^\pm_\sigma$ hoppings~\cite{SuppMat}: 
\begin{equation}
t^\pm_a\neq e^{\pm i \theta} t^\pm_b,
\label{Cond3}
\end{equation}
$\theta$ an arbitrary phase. In other words, the appearance of interaction-induced PBC skin states require \emph{both} (i) unbalanced hoppings for at least one species i.e. usual (non-interacting) NHSE for it had OBCs been implemented, and (ii) the satisfaction of Eq.~\ref{Cond3}, either by having unequal hopping probabilities for each species, or by having complex hoppings not connected by the phase $\theta$ [Fig.~\ref{fig:skin2}b]. 

Physically, PBC skin states around $x_a=x_b$ represent the clusterings of the bosons next to each other. From Eq.~\ref{Cond3}, they occur precisely when the effective bosonic hoppings do not match in terms of probabilities ($|t_a^\pm|\neq|t_b^\pm|$) or phase. As the bosons experience dissimilar directed amplification, they will approach each other on the PBC ring. However, since the interactions prevent double occupancy, they must accumulate near each other, resulting in PBC skin states [Fig.~\ref{fig:skin2}b]. These interaction-induced skin states are only universally observed under PBCs, since under OBCs, particles may also accumulate at the boundaries instead of against each other.

The NHSE origin of these clustered states is substantiated by the drastic changes in the PBC spectrum as the interactions (effective OBCs) are turned on/off. This is demonstrated numerically in [Fig.~\ref{fig:skin2}c], and derived analytically below. From Eq.~\ref{H1D2}, the spectrum in the non-interacting case is simply $\sum_\pm T_\perp^\pm e^{\pm i k_\perp}$, where $\bold k=(k_\text{CM},k_\perp)$ ranges over $[0,2\pi]^2$. However, for interacting cases subject to the PBC skin effect, the ``OBC'' spectrum should be taken over the generalized Brillouin zone (GBZ)~\cite{yao2018edge,lee2019anatomy,yokomizo2019non,lee2018tidal,song2019non,yang2019auxiliary,PhysRevLett.124.086801}~\footnote{In general, skin states and their spectra can be derived via the so-called GBZ construction, which is explained in the Supplement~\cite{SuppMat}.} of $k_\perp$, which for the Hatano-Nelson model is well-known~\cite{SuppMat} to be
\begin{eqnarray}
\bar E^{N=2}&=&2\lambda \sqrt{T^+_\perp T^-_\perp}\notag\\
&=&2\lambda \sqrt{t_a^+t_a^-+t_b^+t_b^-+t_a^+t_b^+e^{2ik_\text{CM}}+t_a^-t_b^-e^{-2ik_\text{CM}}},\notag\\
\end{eqnarray}
where $\lambda$ ranges from $-1$ to $1$. 
Indeed, for unequal unbalanced hoppings (PBC skin effect), this gives a completely different spectrum from the non-interacting case [Fig.~\ref{fig:skin2}c]. While ordinary boundary-induced NHSE is said to break bulk-boundary correspondences, our interaction-induced NHSE breaks the adiabaticity between the interacting/non-interacting limits through particle-particle impenetrability. 

\noindent\textit{Emergent amplification and route towards the thermodynamic limit.--} While the 2-body discussion above admits a complete and intuitive solution, it is in the many-body case that PBC skin states differs qualitatively from ordinary OBC skin states. One new phenomenon is the emergence of continuous amplification in a nearest-neighbor (NN) hopping chain, which never occurs in the non-interacting OBC case i.e. the ``ordindary'' Hatano-Nelson solution. Intuitively, directed amplification under OBCs has to stop after a finite time when a state reaches the fixed boundary, forcing the spectrum to be real. But interaction-induced ``boundaries'' under a PBC setting are dynamic, depending on the positions of all the particles. While two particles always approach the steady state of simply being next to each other, three or more dissimilar particles may approach complicated limit cycles, with say two particles accumulating towards each other at first, and then being repelled by the third, etc. But despite the complexity of such many-body dynamics, they can always be reformulated as a multi-dimensional NHSE problem, whose complicated (but tractable) GBZ encodes the interplay of the various interaction channels.

\begin{figure}
\includegraphics[width=1\linewidth]{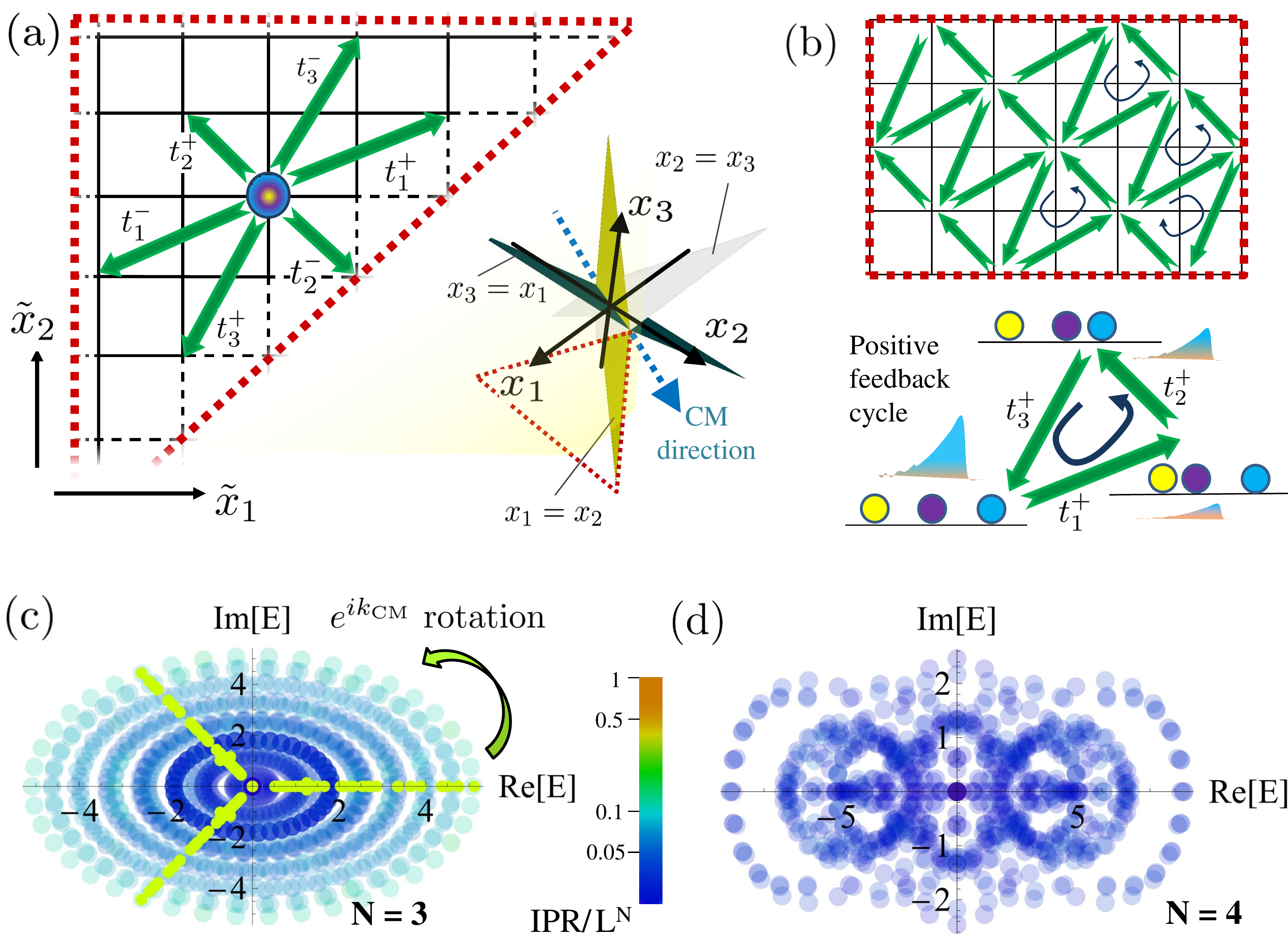}
\caption{(a) PBC skin states accumulate against the $x_1=x_2$, $x_2=x_3$ and $x_3=x_1$ ``boundaries'' of the $N=3$-body configuration space, which are all orthogonal to the $(1,1,1)^T$ CM translation vector. NN hoppings $t_j^\pm$, $j=1,2,3$ on the physical lattice become non-orthogonal and more non-local in the rotated 2D subspace spanned by the ``boundaries''. (b) Effective hoppings in the 2D subspace generate a network of nontrivial loops that amplifies states through a positive feedback cycle. (c) The spectrum of the $t_j^-=0$, $t_j^+=j$ case of Eq.~\ref{poly12main}, with localized skin states (blue) traced out from a Y-shaped locus (schematically colored green) via a $k_\text{CM}$ phase rotation. (d) More complicated spectrum of a $N=4$ case with balanced hoppings $t^\pm_j=j$ for $j=1,2,3$, and unbalanced $t^-_4=4$, $t^+_4=0$, to be contrasted with its non-interacting case~\cite{SuppMat}.
 }
\label{fig:skin3}
\end{figure}
A key insight motivating this reformulation is that interactions that induce PBC skin states conserve the CM. Accumulation in the $k_\text{CM}$ channel requires a physical open boundary, so an $N$-particle PBC skin effect problem described by $H_\text{1D}$ can possess at most $N-1$ independent accumulation channels. Expressing the system in terms of these channels and their symmetries is key to disentangling the many-body dynamics. To be explicit,  $H_{1D}^{N>2}$ possess a $N$-dimensional toroidal configuration space $(x_1,...,x_N)$ partitioned into $(N-1)!$ disconnected regions by $\binom{N}{2}$ ``boundaries'' given by $N_\alpha=N_\beta$, $\alpha,\beta\in {1,...,N}$. For interactions that prohibit more than $n=1$ particles on any site, the $N$ adjacent pairs of particles on the 1D PBC ring demarcate $N$ unique ``boundaries'' for each accessible region. As shown in Fig.~\ref{fig:skin3}a for $N=3$, the two disconnected regions are given by $x_1<x_2<x_3$ and $x_1<x_3<x_2$ modulo cyclic permutations, and are separated by the ``boundaries'' $x_1=x_2$, $x_2=x_3$ and $x_3=x_1$. Importantly, the normals $(1,-1,0)^T,(0,1,-1)^T$ and $(-1,0,1)^T$ to all these ``boundaries'' are all orthogonal to the CM translation direction $(1,1,1)^T$. To find the PBC skin states, we must thus construct the multi-dimensional GBZ~\cite{Lee2019hybrid} in a rotated basis orthogonal to the CM momentum $k_\text{CM}$, as detailed in~\cite{SuppMat} for generic number of particles $N$.

For $N=3$ particles, our Hamiltonian can be re-written in terms of rotated momenta $\tilde{\bold k}=(\tilde k_1,\tilde k_2,k_\text{CM})$ as
\begin{eqnarray}
H_\text{1D}^{N=3}(\bold k)&=& \sum_\pm t^\pm_1 e^{\pm ik_1}+t^\pm_2 e^{\pm ik_2}+t^\pm_3 e^{\pm ik_3}\notag\\
&=& \sum_\pm t^\pm_1 e^{\pm i(k_\text{CM}+2\tilde k_1+\tilde k_2)}+t^\pm_2 e^{\pm i(k_\text{CM}-\tilde k_1+\tilde k_2)}\notag\\
&&+t^\pm_3 e^{\pm i(k_\text{CM}-\tilde k_1-2\tilde k_2)},
\label{poly0main}
\end{eqnarray}
where $t_j^\pm$ are the right/left hoppings amplitudes of particle $j$, equivalent across different particles $j$ of the same species [Fig.~\ref{fig:skin3}a]. One easily checks that the combined amplitude of hopping particles $1$ and $2$ towards each other is $t_1^\pm t_2^\mp e^{\pm i(k_1-k_2)}=t_1^\pm t_2^\mp e^{\pm 3i\tilde k_1}$, which depends solely on $\tilde k_1$. Thus the skin states corresponding to their repulsion can be encoded by the GBZ of $\tilde k_1$ alone. Likewise, the skin states due to the repulsion between particles $2$ and $3$ is encoded by the GBZ of $\tilde k_2$ alone. 

This basis rotation for obtaining PBC skin states is not just a notational change, but has profound dynamical consequences in fact. In Fig.~\ref{fig:skin3}b, shown for only right hoppings $t_j^+$ with $N=3$, hoppings that are originally across NN physical lattice sites now possess multiple hopping ranges in each direction. In particular, they now form a network with nontrivial directed loops, which are dynamically nontrivial since asymmetric hoppings lead to amplification or attenuation. For instance, a state can now cycle through a series of directed hoppings $t_1^+\rightarrow t_2^+\rightarrow t_3^+\rightarrow t_1^+$ (or $t_1^+\rightarrow t_3^+\rightarrow t_2^+\rightarrow t_1^+$) back to its original configuration, and experience net amplification. Physically, such cycles correspond to successive shifts of the individual particles, each incurring skin state accumulation due to the hopping asymmetry, such that the entire particle configuration collective translates in accordance to the CM wavevector $k_\text{CM}$.  

Due to these perpetual amplification cycles, complex eigenenergies emerge in the interacting spectrum for $N>2$, even though the single-particle spectra remains real. Through iterated GBZ constructions in the rotated basis, where effective hoppings can extend non-locally up to $N-1$ sites~\cite{SuppMat}, one can derive the full complex PBC skin state spectrum $\bm\bar{E}^{N}|_{t^-_j=0}$ for arbitrarily many particles $N$ experiencing only rightwards NN hoppings $t^+_\sigma$:
\begin{eqnarray}
\bm\bar{E}^{N}|_{t^-_j=0}\propto e^{ik_\text{CM}}\prod_{\sigma=1}^st_\sigma^{f_\sigma}e^{2\pi i \nu/N},
\label{poly12main}
\end{eqnarray}
where $f_\sigma$ are the fractional populations of species $\sigma=1,...,s$. $\bm\bar{E}^{N}|_{t^-_j=0}$ forms a star-shaped locus with $N$ ``spikes'' $\nu=1,...,N$ in the complex energy plane, $\nu$ labeling the $N$ possible sectors for cyclical amplification [Fig.~\ref{fig:skin3}c for $N=3$]. For each cycle of duration $\Delta t$, the amplification factor is bounded above by $e^{\text{Im}\,\bar{\bar E}\Delta t}\sim e^{|\sin(k_\text{CM}+2\pi \nu/N)|\Delta t}$, which is trivial only if $2\pi\nu/N$ cancels the CM density wave vector $k_\text{CM}$. Shown in Fig.~\ref{fig:skin3}b, for instance, is the $\nu=0$ sector with $k_\text{CM}=0$, with the 3-particle state translating into itself. Although $k_\text{CM}$ did not explicitly enter the Hamiltonian $H_\text{1D}$, it modulates the propensity of gain/loss by indirectly restricting the interference of accumulated skin states, as suggested by Eq.~\ref{poly12main}. For larger $N$, more exotic complex spectra [Fig.~\ref{fig:skin3}d] and dynamical behavior can be similarly computed and contrasted with their non-interacting counterparts~\cite{SuppMat}. In the thermodynamic limit, the spectrum generically depends only on the fractional populations of the species and their hoppings.

\begin{figure}
\includegraphics[width=1\linewidth]{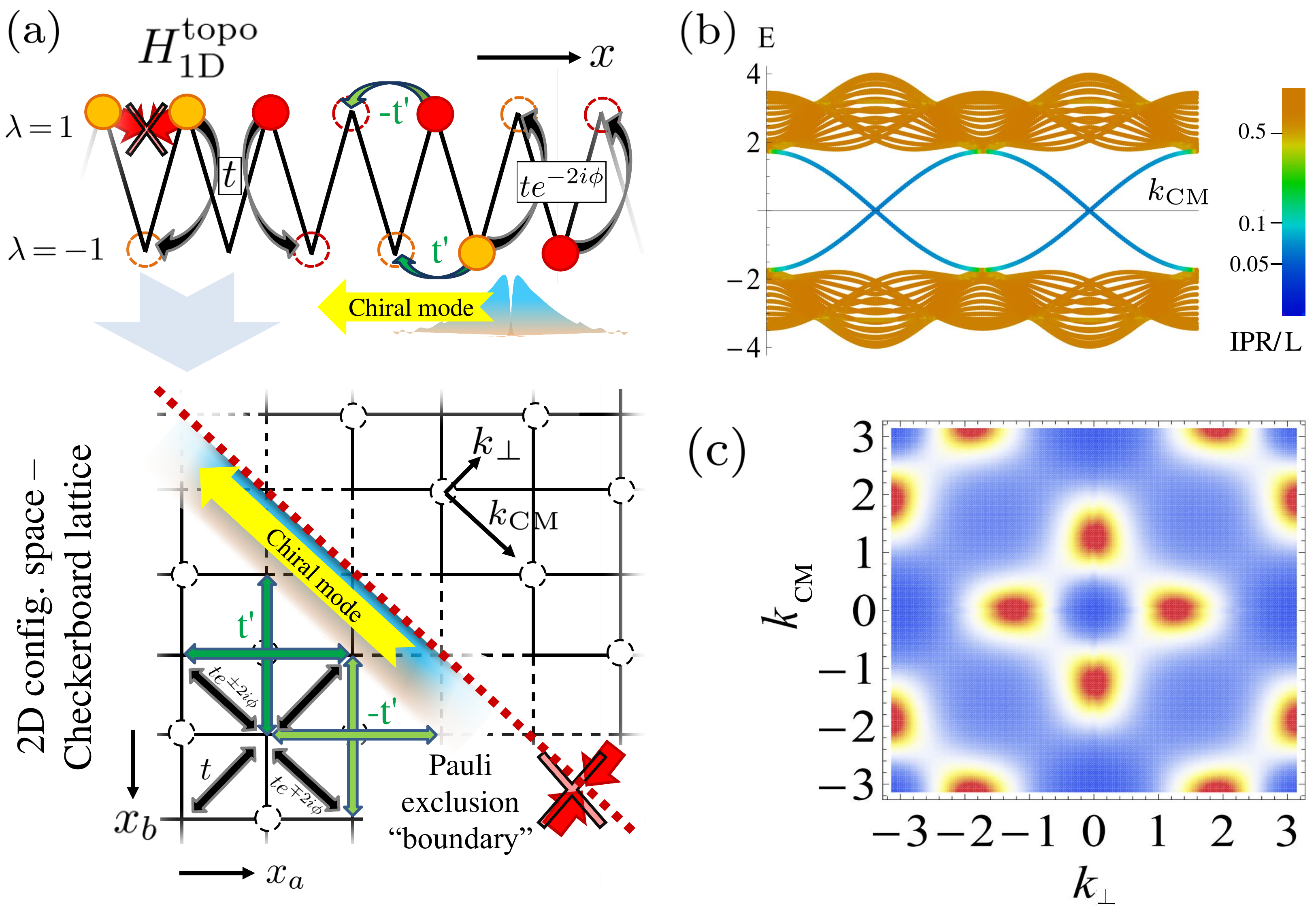}
\caption{(a) Schematic of our PBC zigzag chain with 1 and 2-fermion hoppings $\pm t'$ and $t$ [Eq.~\ref{Htopo}], with a flux $\pm 2\phi$ experienced by two fermions hopping from level $\lambda =\pm 1$ to $\mp 1$ in the same $x$-direction. It maps to an OBC Checkerboard lattice with topologically protected simultaneous two-body chiral propagation. (b) Energy bandstructure of $H^\text{topo}_\text{1D}$ for $t=t'=1$ and $\phi=\pi/6$, marked by $2$ localized in-gap modes. (c) Its Berry curvature integrating to a Chern number of $2$. 
}
\label{fig:topo}
\end{figure}

\noindent\textit{Chiral topological modes from many-body effects.--} In PBC lattices with non-trivial unit cells, many-body can also induce topologically protected states that normally exist only at open boundaries. These states are protected by topological invariants of the many-body configuration space bulk, and similarly appear along the ``boundaries'' where particles coincide. An illustrative PBC lattice with Chern~\cite{haldane1988model,qi2008} ``boundary modes'' is given by a two-level 1D zigzag chain containing 2 fermions $\mu,\nu$ that have to be simultaneously on the same level $\lambda=(-1)^x$ at any time [Fig.~\ref{fig:topo}a]. The simplest and most local ways they can hop are: (i) one-body level and species-dependent hoppings $\pm t'$ across two sites, and (ii) simultaneous two-body hoppings $t$ to adjacent sites on the other level. To break time reversal symmetry, a flux drives $t\rightarrow t\, e^{\pm 2\lambda i\phi}$ for fermions who simultaneously hop in the same $x$ direction. Their Hamiltonian is thus given by
\begin{eqnarray}
H_\text{1D}^\text{topo}&=&t\sum_{\Delta x_1,\Delta x_2=\pm 1}\mu^\dagger_{x_1+\Delta x_1}\nu^\dagger_{x_2+\Delta x_2}\mu_{x_1}\nu_{x_2}e^{i\lambda\phi|\Delta x_1+\Delta x_2|}\notag\\
&&+t'\,\sum_{x}\lambda (\mu^\dagger_{x+2}\mu_{x}-\nu^\dagger_{x+2}\nu_{x})+\text{h.c.}
\label{Htopo}
\end{eqnarray}
In its 2-body configuration space, $H_\text{1D}^\text{topo}$ maps onto a 2D checkerboard lattice Hamiltonian with Chern number $C=\pm 2$ bands, related to that of Refs.~\cite{sun2011nearly,lee2014lattice,regnault2011fractional} and detailed in~\cite{SuppMat}. With Pauli exclusion demarcating a ``boundary'' along $x_1=x_2$, the system is translation invariant only along the CM direction, and exhibits $C=2$ in-gap topological modes~\footnote{This is the Chern number in the 2D configuration space, unlike Ref.~\cite{yoshida2019non} which computed the many-body Chern number in a 2D physical system.} [Fig.~\ref{fig:topo}](b-c). Physically, these ``boundary'' chiral modes represent correlated fermion clusters that move with nontrivial CM group velocity due to the asymmetry from the fluxes and occupancy-dependent hopping probabilities~\cite{SuppMat}.

\noindent\textit{Discussion.--} Particle statistics and appropriate interactions can introduce effective configuration space ``boundaries'' that support bona fide skin or topological states. Compared to other more common repulsive mechanisms based on energetics or particle statistics, our hopping attenuation mechanism allows for more targeted engineering of robust PBC states. However, the dynamical nature of these ``boundaries'' qualitatively modifies the effective hoppings, leading to possible instabilities. Such cases with complex spectra, however, may still possess stable Rabi oscillations dynamics between similarly divergent states~\cite{lee2020ultrafast}. Away from fine-tuning, interactions will generically lead to partial ``boundaries'' that exhibit weaker albeit still robust skin/topological localization~\cite{SuppMat}. While PBC skin states generalize straightforwardly in the thermodynamic (large $N$) limit, protection of their topological counterparts rely on high-dimensional topological invariants~\cite{bernevig2003eight,prodan2013non,hasebe2014higher,lee2018electromagnetic,petrides2018six} that will be interesting for future studies.




\bibliography{references}

\clearpage

\onecolumngrid
\noindent\textbf{\large Supplemental Online Material for ``Many-Body Topological and Skin States without Open Boundaries" }\\[5pt]
\vspace{0.1cm}
{\small This supplementary contains the following material arranged by sections:\\
\begin{enumerate}
\item Brief background about non-Hermitian skin spectra, detailed treatment of 2-body and selected 3-body and $N$-body cases. The general basis transformation for decoupling the center of mass (CM) degree of freedom and the emergence of non-local hoppings that give rise to very different dynamical properties. Discussion of non fine-tuned interactions that only partially restrict the Hilbert space.
\item Details of the mapping between $H_\text{1D}^\text{topo}$ of the main text and a Checkerboard model with nontrivial Chern number. Analysis of its chiral ``boundary'' modes.
\end{enumerate}
}
\setcounter{equation}{0}
\setcounter{figure}{0}
\setcounter{table}{0}
\setcounter{page}{1}
\setcounter{section}{0}
\setcounter{secnumdepth}{3}
\makeatletter
\renewcommand{\theequation}{S\arabic{equation}}
\renewcommand{\thefigure}{S\arabic{figure}}
\renewcommand{\thesection}{S\Roman{section}}
\renewcommand{\thepage}{S\arabic{page}}
\vspace{1cm}

\section{I. Skin spectra for many-particle PBC systems}

\subsection{Background}

In our work, a many-body system with unequal hoppings felt by different particles is mapped to a multi-dimensional configuration space lattice with unbalanced gain/loss. Given such a lattice Hamiltonian with left and right hoppings of unbalanced magnitudes, we expect to observe a robust accumulation of particles whenever there is lattice inhomogeneity. This is known as the non-Hermitian skin effect (NHSE), and is characterized by large, non-perturbative changes in the spectrum and eigenstate distribution by the inhomogeneities. It can be rigorously treated in the so-called generalized brillouin zone (GBZ), where an imaginary part is added to the lattice momentum to obtain a surrogate Hamiltonian which no longer exhibits the NHSE~\cite{lee2019unraveling,lee2019anatomy,yokomizo2019non,lee2018tidal,yang2019auxiliary,PhysRevLett.124.086801}, at least when away from criticality~\cite{li2020critical,liu2020helical,lee2020ultrafast}. While most other works have focused on open boundaries as the source of the lattice inhomogeneities, in this work we instead consider interactions that penalizes configurations with multiple occupancy as the origin of ``boundaries'' in the many-body configuration space lattice.

There are already several excellent treatments on the NHSE, as referenced in the main text, so here we shall only summarize the key steps, as well as review a few key results necessary analyzing our so-called PBC skin effect in the many-body configuration space.

\begin{enumerate}
\item For sufficiently simple multi-dimensional lattice in a generic configuration space, the NHSE can be analyzed iteratively, one orthogonal direction at a time. Treating the transverse momentum components as external parameters, the lattice is described as a quasi-1D Hamiltonian $H_\text{1D}(k_\perp)$, where $k_\perp$ is normal to the ``boundary'' of interest.
\item To restore the ``bulk boundary correspondence'', which in our context allows us to understand the effects of the interactions, the NHSE can be ``gauged away'' by analytically continuing $k_\perp$ in the complex plane i.e. constructing the GBZ. This is done via $k_\perp\rightarrow p_\perp + i\kappa(p_\perp)$, where $p_\perp \in \mathbb{R}$ and $\kappa(p_\perp)$ is the smallest complex deformation such that there exists another eigensolution with the \emph{same} eigenenergy and $\kappa$. This is purely an algebraic property of the characteristic polynomial of the energy eigenequation. In Hermitian or reciprocal systems, $\kappa(p_\perp)=0$ by construction.
\item In the basis of the GBZ, the surrogate Hamiltonian~\cite{lee2019unraveling} $\bar H_\text{1D}(p_\perp)=H_\text{1D}(p_\perp+i\kappa(p_\perp))$ no longer experiences the NHSE, and its spectrum exhibits adiabatic continuity as lattice inhomogeneities are introduced. Sometimes, this is also written as $z\rightarrow z\,e^{-\kappa(p_\perp)}$, where $z=e^{ip_\perp}$. The corresponding eigenenergies of the skin states are denoted as $\bar E$ or, after two GBZ constructions, as $\bar{\bar E}$ etc.

Physically, the skin state accumulation is characterized by the inverse decay length/skin depth $\kappa(p_\perp)$, which can be different for different wavenumbers $2\pi/p_\perp$ of its oscillatory part.

After performing the GBZ construction for all directions, we should arrive at a surrogate Hamiltonian depending on real inverse wave numbers $\bold p$ that does not exhibit the NHSE, but whose spectrum gives the correct skin state eigenenergies.
\end{enumerate}

\subsubsection{Analytic results for archetypal scenarios}

In our context, the equivalent multi-dimensional NHSE system corresponding to the physical multi-particle interacting system often takes the effective form 
\begin{equation}
E=Az^\alpha+Bz^{-\beta},
\label{EAB0}
\end{equation}
where $E$ is the eigenenergy and $z=e^{ik_\perp}$ describes the momentum component normal to the ``boundary'' of interest. This effective quasi-1D system has only two hoppings: one with amplitude $A$ and $\alpha$ sites to the right, and the other with amplitude $B$ $\beta$ sites to the left. While most literature have only considered left/right imbalances with $A\neq B$ but $\alpha=\beta=1$, solving our many-body system with CM conservation also requires understanding of what happens with dissimilar hopping ranges ($\alpha\neq \beta$), as explained in the main text.

The $\alpha=\beta=1$ case $E=Az+B/z$ is known as the Hatano-Nelson model~\cite{HN1996prl}. Under OBCs in the effective lattice (albeit still with PBCs in the physical lattice of the systems we consider),  its skin modes possess the simple solution
\begin{equation}
\bar E = 2\sqrt{AB}\cos p
\label{barE}
\end{equation}
where $p=p_\perp +\text{Arg}\left[\sqrt{A/B}\right]$ ranges from $0$ to $2\pi$. As such, the skin states energies linearly interpolate between $-2\sqrt{AB}$ to $2\sqrt{AB}$, and is completely real if $\sqrt{AB}$ is real. The skin depth is independent of $p_\perp$, and is given by
\begin{equation}
\kappa(p_\perp)=\kappa=\frac1{2}\log\left|\frac{A}{B}\right|.
\end{equation}
Indeed, $\bar E$ can be obtained from $E=Az+B/z$ via the GBZ defined by $z\rightarrow e^{-\kappa}e^{ip_\perp}=\sqrt{\left|\frac{B}{A}\right|}e^{ip_\perp}$.

More generally~\cite{lee2019unraveling}, the $\alpha,\beta \neq 1$ case admits skin state solutions that trace star-shaped loci in the complex energy plane, i.e.
\begin{equation}
\bar E \propto A^{\frac{\beta}{\alpha+\beta}}B^{\frac{\alpha}{\alpha+\beta}}\omega_{\alpha+\beta}^\nu,
\label{barE2}
\end{equation} 
where $\omega_{\alpha+\beta}$ is a $\alpha+\beta$-th root of unity and $\nu=1,...,\alpha+\beta$. As $p$ ranges over $0$ to $2\pi$, the eigenenergies sweep through the ``spikes'' of the $\alpha+\beta$-pronged star successively (For illustration, the $\cos p$ factor in Eq.~\ref{barE} describes the back-and-forth sweep on a 2-pointed star i.e. line segment).

\subsection{Analytic results for the 2-body case}

We consider the Hamiltonian $H_\text{1D}$ (Eq.~1 and 2 of the main text) with $n=1$ interaction $V^\pm_\sigma = t^\pm_\sigma$ that prohibits double occupancy, with two bosons of species $a,b$ respectively:
\begin{eqnarray}
H_\text{1D}^{N=2}&=& \frac1{2}\sum_{\pm,i}\sum_{\sigma\in \{a,b\}}t^\pm_\sigma c^\dagger_{i\pm 1,\sigma}c_{i,\sigma}(2-\rho_{i\pm 1})\notag\\
&=&\sum_{\bold k,\pm}T^\pm_\perp\, e^{\pm i k_\perp}
\end{eqnarray}
where $T^\pm_\perp=t_a^\pm e^{\pm i k_\text{CM}}+t_b^\mp e^{\mp i k_\text{CM}}$. For PBC skin states to emerge due to particle clustering when interactions are switched on, we require $|T_\perp^+|\neq|T_\perp^-|$, at least for some $k_{CM}$. Squaring both sides and separately considering coefficients of different powers of $e^{\pm i k_\text{CM}}$, we find that interaction-induced PBC skin states will be \emph{absent} only when the both conditions
\begin{subequations}
\begin{equation}
\frac{t_a^+}{t_b^+}=\left(\frac{t_a^-}{t_b^-}\right)^*,
\label{cond1}
\end{equation}
\begin{equation}
|t^+_a|^2-|t^-_a|^2= |t^+_b|^2-|t^-_b|^2
\label{cond2}
\end{equation}
\end{subequations}
are satisfied. 
Eq.~\ref{cond1} states that both species must possess skin states of identical inverse decay lengths (skin depths) $\log\frac{|t_\sigma^+|}{|t_\sigma^-|}$ in physical real space under OBCs, at the non-interacting level. Combined with Eq.~\ref{cond2}, we further require that either (i) there is no physical non-reciprocity at all i.e. $|t_\sigma^+|=|t_\sigma^-|$ for $\sigma=a,b$, or that (ii) $t^\pm_a/t^\pm_b=e^{\pm i\theta}$, $\theta$ an arbitrary phase. But (ii) subsumes the implication of Eq.~\ref{cond1} on identical skin depths. Hence, we conclude that interaction-induced PBC skin states are \emph{absent} whenever the two species $a,b$ independently experience hoppings related by
\begin{equation}
t^\pm_a =e^{\pm i \theta} t^\pm_b,
\label{cond3}
\end{equation}
even though each may possess identically decaying skin states under OBCs. Note, however, that the converse is not necessarily true: Even if each species possess OBC skin states of equal decay lengths, i.e. $\frac{|t_a^+|}{|t_a^-|}=\frac{|t_b^+|}{|t_b^-|}$, they can still exhibit interaction-induced PBC skin states if the phases of the hoppings do not obey Eq.~\ref{cond3}.

The conditions are more relaxed when we desire only to not have PBC skin states satisfying specific criteria. For instance, the absence of translation invariant PBC skin states i.e. those in the $k_\text{CM}=0$ sector only requires that $|t_a^++t_b^-|=|t_b^++t_a^-|$. Also, the absence of PBC skin states that depend on $k_\text{CM}$ requires only that $t_a^+t_a^-=t_b^+t_b^-$.

\begin{figure}
\subfloat[]{\includegraphics[width=.4\linewidth]{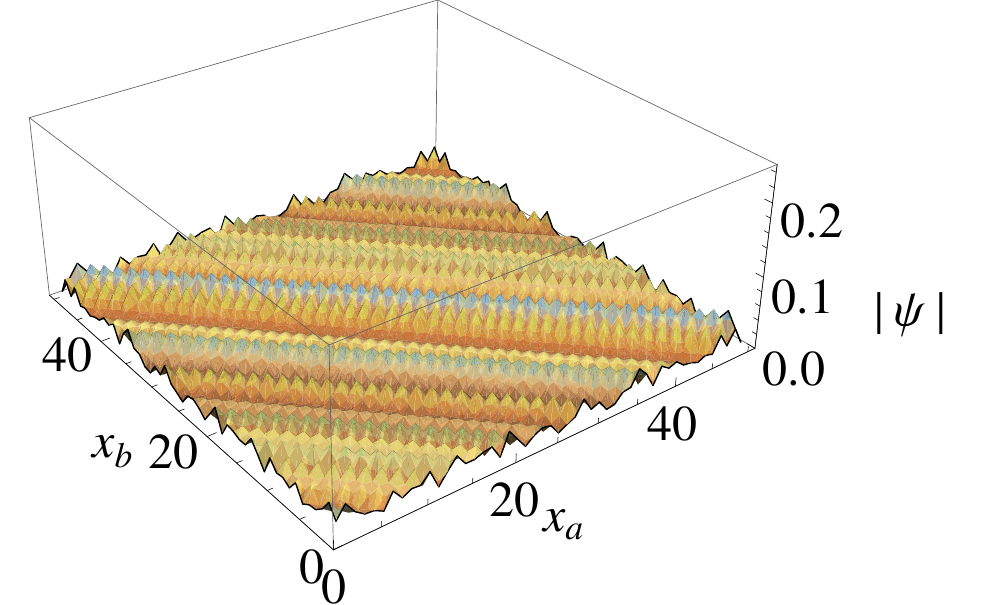}}
\subfloat[]{\includegraphics[width=.4\linewidth]{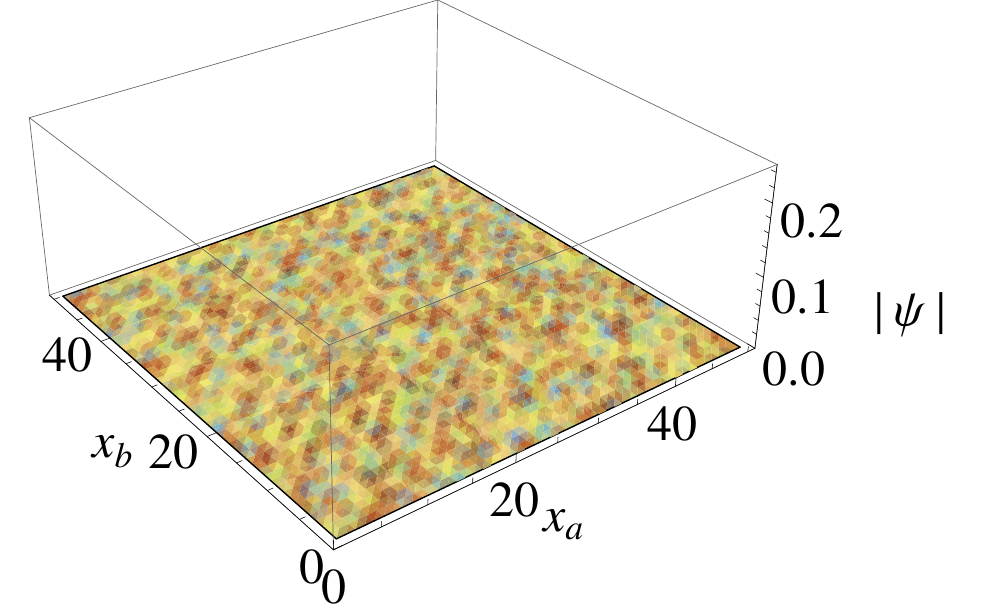}}\\
\subfloat[]{\includegraphics[width=.4\linewidth]{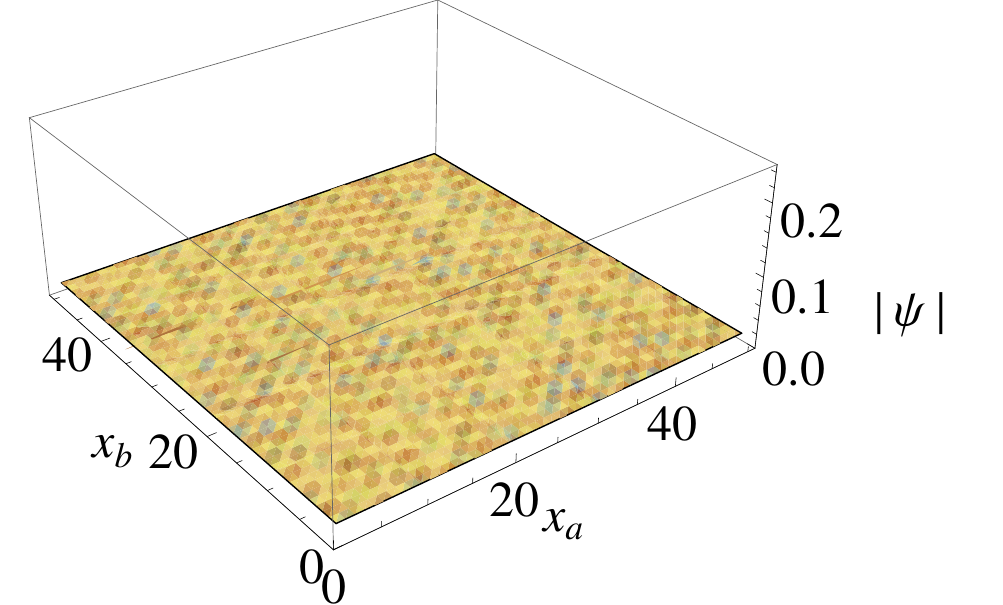}}
\subfloat[]{\includegraphics[width=.4\linewidth]{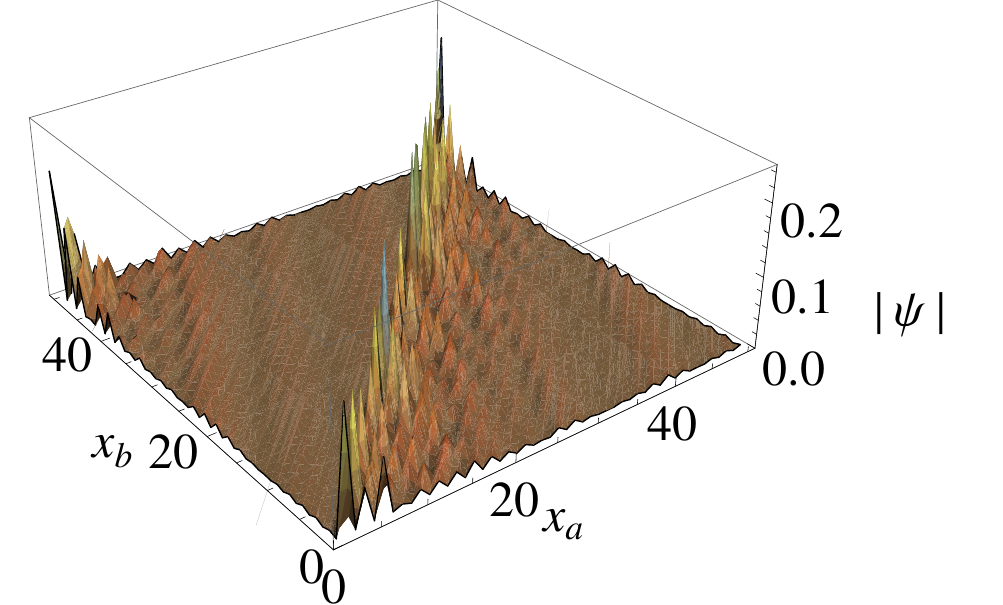}}\\
\subfloat[]{\includegraphics[width=.4\linewidth]{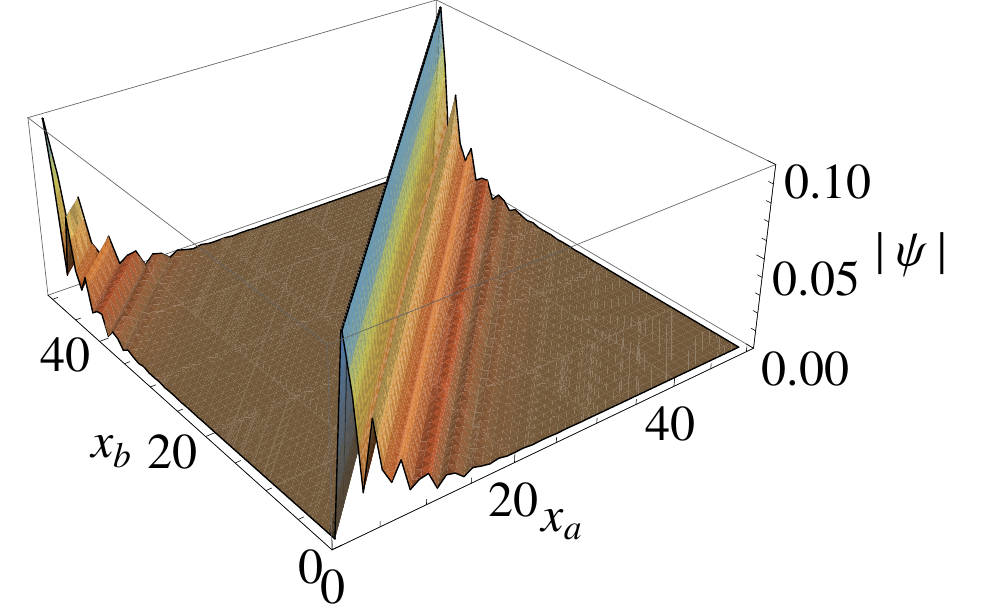}}
\subfloat[]{\includegraphics[width=.4\linewidth]{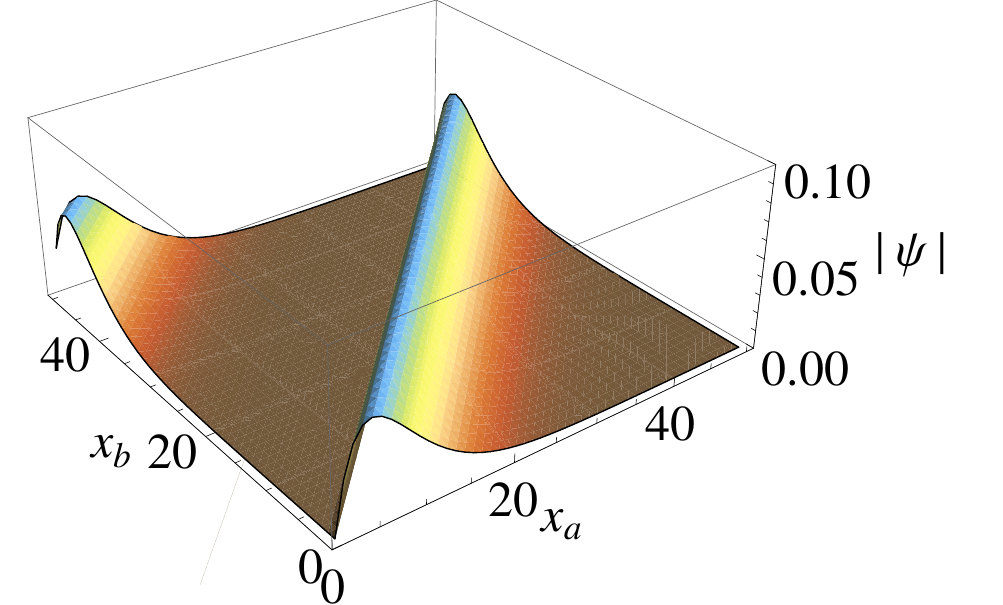}}
\caption{Profile of selected eigenstates of $H_\text{1D}$ for $N=2$ bosons. (a,b) Representative eigenstates at $E=0$ for the identical unbalanced hoppings case of Fig.~1 of the main text, with interactions $V^\pm_\sigma$ turned off and on respectively. There are no skin states and hence no qualitative difference is observed. (c,d) Representative eigenstates at $E=0$ for the identical unbalanced hoppings case of Fig.~1 of the main text, also with interactions $V^\pm_\sigma$ turned off and on respectively. Evidently, interactions are necessary for spatial skin localization (blue) along $x_a=x_b$. (e,f) PBC skin states at other energies for the same system, at $E=-3$ and $E=-6.6+0.27i$ respectively. In fact, skin localization is pervasive across all eigenstates, once it appears. }
\label{fig:N=2}
\end{figure}

\subsection{Treatment for many-body cases}

We consider having $N>2$ particles, but still with interactions of $n=1$ i.e $t^\pm_\sigma = V^\pm_\sigma$ for simplicity, such that no double occupancy is allowed. In this case, the configuration space of $H^{N>2}_\text{1D}$ is a N-dimensional torus with $\binom{N}{2}$ ``boundaries'' defined by disallowed double occupancy configurations i.e. $x_{\alpha_1}=x_{\alpha_2}$ where $\alpha_1,\alpha_2\in{1,...,N}$ label the bosons. These ``boundaries'' partition the configuration space into $(N-1)!$ disconnected regions, each corresponding to a configuration $x_{\alpha_1}\ll x_{\alpha_2}\ll\dots\ll x_{\alpha_N}$ and cyclic permutations since the physical system is periodic (Had it been OBCs, we would have obtained $N!$ disconnected regions). For instance, with $N=3$ we have $\binom{3}{2}=3$ ``boundaries'' $x_1=x_2$, $x_1=x_3$ and $x_2=x_3$. The $(3-1)!=2$ disconnected regions are namely the configurations with $x_1<x_2<x_3$ and cyclic permutations, as well as the configurations with $x_1<x_3<x_2$ and cyclic permutations.

Each disconnected region is bounded by the $N$ ``boundaries'' $x_{\alpha_1}=x_{\alpha_2}$, $x_{\alpha_2}=x_{\alpha_3}$, ..., $x_{\alpha_N}=x_{\alpha_1}$. Generically, the NHSE will accumulate skin states against all of these boundary surfaces. However, these ``boundaries'' collectively form an $N-1$-dimensional subspace, even though there are $N$ of them. This is because their normals all take the form $\hat e^{\alpha_i}-e^{\alpha_j}$, and are orthogonal to the direction of center-of-mass translation $\hat e_1+\hat e_2+...+\hat e_N$. As such, the skin state accumulation occurs only in $N-1$ independent directions, with the center-of-mass momentum $k_\text{CM}=\frac1{N}\sum_{i=1}^Nk_i$ \emph{not} to be complex-deformed in the multidimensional generalized BZ. (otherwise, that presupposes that skin state accummulation also partakes in that sector, which is untrue).

To avoid deforming the momentum components containing $k_\text{CM}$, we introduce a basis transformation $M$ connecting the orthonormal configuration space lattice basis $\hat e_j$ with the basis spanned by the normals of the $x_{\alpha_j}=x_{\alpha_{j+1}}$ ``boundaries'':
\begin{equation}
\Delta \tilde x = M\Delta x,
\end{equation}
where the $j$-th component of $\Delta x$ represent the displacement of the $j$-th particle, while the $j$-th component of $\Delta \tilde x$ represent the relative displacement of the $j$-th and $j+1$-th particles (such that it disappears if the $j$-th and $j+1$-th particles collide). The $N$-th (final) component of $\Delta \tilde x$ represents the center-of-mass displacement $\Delta x_\text{CM}=\frac1{N}\sum_{i=1}^N \Delta x_i$, so that we have $\tilde \Delta x = (\Delta \tilde x_1,...,\Delta \tilde x_{N-1},\Delta x_\text{CM})$. The conjugate momentum components are defined analogously:
\begin{equation}
 \tilde k^T = k^T M^{-1},
\end{equation}
with the components of $k$ and $\tilde k$ being the momenta dual to $x$ and $\tilde x$.  
$\tilde k_N$ is defined to be $k_\text{CM}=\frac1{N}\sum_{i=1}^N k_i$, so that $\tilde k = ( \tilde k_1,...,\tilde k_{N-1},k_\text{CM})$. As required, the scalar product $k^T\cdot \Delta x = \tilde k^T\cdot \Delta \tilde x$ remains invariant. 

Explicitly, $M^{-1}$ is an $N\times N$ matrix taking the form
\begin{equation}
M^{-1}=\frac1{N}\left(\begin{matrix}
1 & 0 & 0 &...& 0 & 1 \\
-1 & 1 & 0 & ...& 0 & 1 \\
0 & -1 & 1 & ... & 0 & 1 \\
0 & 0 & -1 & ... & 0 & 1 \\
\vdots & \vdots & \vdots & \ddots & \vdots & \vdots \\
0 & 0 & 0 & ... & 1 & 1 \\
0 & 0 & 0 & ... & -1 & 1 \\
\end{matrix}\right), 
\label{Minv}
\end{equation}
such that $M^{-1}(1,0,0,...)^T=(1,-1,0,...)^T$ gives the normal to the $x_1=x_2$ ``boundary'', $M^{-1}(0,1,0,...)^T=(0,1,-1,...)^T$ gives the normal to the $x_2=x_3$ ``boundary'' etc. Note that for $N\geq 4$, each configuration space region is bounded by only $N$ out of the $\binom{N}{2}$ ``boundaries'', and the ``1'' and ``-1''s in the columns in $M^{-1}$ should be placed according to the normals of the specific ``boundaries'' present in the particular disconnected region of interest.

From Eq.~\ref{Minv}, $M$ takes the nice form
\begin{equation}
M=\left(\begin{matrix}
N-1 & -1 & -1 & -1 &... &-1& -1\\
N-2 & N-2 & -2 & -2 &... &-2& -2 \\
N-3 & N-3 & N-3 & -3 & ...&-3 & -3 \\
\vdots & \vdots & \vdots & \vdots & \ddots & \vdots & \vdots \\
2 & 2 & 2 & 2 & ...&-N+2  & -N+2 \\
1 & 1 & 1 & 1 & ... & 1 & -N+1 \\
1 & 1 & 1 & 1 &...& 1  & 1 \\
\end{matrix}\right).
\label{M} 
\end{equation}
Importantly, by writing $k=M^T\tilde k$, we see that if the only nonzero $\tilde k$ component is $\tilde k_i$, $i=1,...,N-1$, the $k_j-k_{j'}$ will vanish unless $j=i$, $j'=i+1$, where it will be $N\tilde k$. This means that the $\tilde k_i$ direction indeed represents the direction where particles $i$ and $i+1$ are brought together. This can be more explicitly observed by writing
\begin{equation}
k_j=\sum_{i=1}^N\tilde k_iM_{ij}=k_\text{CM}+\sum_{i=1}^{N-1}\left[N\theta(i-j)-i\right]\tilde k_i
\end{equation}
where the Heaviside function is defined as $\theta(x)=1$ for $x\geq 0$.

An arbitrary many-body hopping amplitude $t$ across $\Delta x_1$ sites for the first particle, $\Delta x_2$ sites for the second particle, etc. takes the momentum-space form $t\,e^{ik^T\cdot\Delta x}$, where $\Delta x =(\Delta x_1,\Delta x_2,...)$ in the $N$-boson configuration space. Rewriting the scalar product as $k^T\cdot\Delta x = k^TM^{-1}M\Delta x= \tilde k^TM\Delta x$, we can express the hopping term in terms of the transformed momentum (but still defined through original physical hopping displacements) as 
\begin{eqnarray}
te^{i\tilde k^T\cdot M\Delta x}&=& te^{i\sum_{i,j=1}^N \tilde k_iM_{ij} \Delta x_j}\notag\\
&=&te^{iN\Delta x_\text{CM}\left(k_\text{CM}-\sum_{j=1}^{N-1}j\tilde k_j\right)}e^{iN\sum_{j=1}^{N-1}\tilde k_j\sum_{i=1}^j \Delta x_i}\notag\\
&=&te^{iN\Delta x_\text{CM}k_\text{CM}}e^{-iN\sum_{j=1}^{N-1}\left(j\Delta x_\text{CM}-\sum_{i=1}^j\Delta x_i\right)\tilde k_j}
\label{hop1}
\end{eqnarray}
where $\Delta x_\text{CM}=\frac1{N}\sum_{j=1}^N\Delta x_j$. Indeed, if $\Delta x$ were to involve only particles $j$ and $j+1$ being brought together, i.e. $\Delta x\propto \hat e_j-\hat e_{j+1}$, we will see that only $\tilde k_j$ appears on the RHS.

If the hopping only involves a single particle $l$, across $\Delta x_l$ sites, $\Delta x_i=\delta_{il}\Delta x_i$ and Eq.~\ref{hop1} simplifies to
\begin{eqnarray}
te^{i\tilde k^T\cdot M\Delta x_l\hat e_l}&=& te^{i\Delta x_l k_\text{CM}}e^{-i\Delta x_l\sum_{j=1}^{N-1}\left(j-N\theta(j-l)\right)\tilde k_j}.
\label{hop2}
\end{eqnarray}
In other words, from Eq.~\ref{M}, the exponents in the hopping of particle $l$ can be read from the columns of $M$, while the coefficients of $\tilde k_j$ across of the hopping exponents of all particles can be read from the rows of $M$.

In both Eqs.~\ref{hop1} and \ref{hop2}, the factor containing $k_\text{CM}$ is not subject to complex deformation for the GBZ construction, and can safely be treated as constants. The other terms are degree $N-1$ multi-variate monomials in $z^j=e^{i\tilde k_j}$, and will combine with other hoppings to form the multi-variate characteristic polynomial which will be used to derive the $N-1$-D GBZ as well as skin spectrum.

\subsubsection{Hoppings with $N=3$ particles}
\noindent For explicit illustration and later application, we list down explicit forms the single-particle hoppings with $N=3$ particles, for the 1st, 2nd and 3rd particle respectively:
\begin{subequations}
\begin{equation}
te^{i\tilde k^T\cdot M\Delta x_l\hat e_1}|_{N=3}=te^{i\Delta x_1 (k_\text{CM}+2\tilde k_1+\tilde k_2)}
\end{equation}
\begin{equation}
te^{i\tilde k^T\cdot M\Delta x_l\hat e_2}|_{N=3}=te^{i\Delta x_2 (k_\text{CM}-\tilde k_1+\tilde k_2)}
\end{equation}
\begin{equation}
te^{i\tilde k^T\cdot M\Delta x_l\hat e_3}|_{N=3}=te^{i\Delta x_3 (k_\text{CM}-\tilde k_1-2\tilde k_2)}.
\end{equation}
\label{hop3}
\end{subequations}
Multi-particle hoppings can be simply written down by multiplying the constituent single-particle hoppings.

\subsubsection{Hoppings with $N=4$ particles}
\noindent For further explicit illustration, we also list down explicit forms the single-particle hoppings with $N=4$ particles, for the 1st, 2nd, 3rd and 4th particle respectively:
\begin{subequations}
\begin{equation}
te^{i\tilde k^T\cdot M\Delta x_l\hat e_1}|_{N=4}=te^{i\Delta x_1 (k_\text{CM}+3\tilde k_1+2\tilde k_2+\tilde k_3)}
\end{equation}
\begin{equation}
te^{i\tilde k^T\cdot M\Delta x_l\hat e_2}|_{N=4}=te^{i\Delta x_2 (k_\text{CM}-\tilde k_1+2\tilde k_2+\tilde k_3)}
\end{equation}
\begin{equation}
te^{i\tilde k^T\cdot M\Delta x_l\hat e_3}|_{N=4}=te^{i\Delta x_3 (k_\text{CM}-\tilde k_1-2\tilde k_2+\tilde k_3)}
\end{equation}
\begin{equation}
te^{i\tilde k^T\cdot M\Delta x_l\hat e_4}|_{N=4}=te^{i\Delta x_4 (k_\text{CM}-\tilde k_1-2\tilde k_2-3\tilde k_3)}.
\end{equation}
\end{subequations}

Note that our basis transformation is not the only valid one; i.e. for interactions with fuller rotational symmetry, Barycentric coordinates, which has been used for constructing generalized quantum Hall pseudopotentials~\cite{lee2015geometric}, may be more appropriate.
In fact, any rotation to an orthogonal basis containing $k_\text{CM}$ will be able to decouple the latter from being coupled with the other degrees of freedom that will be deformed under GBZ construction. 

\subsection{Explicit constructions of PBC skin states and spectra for $N=3$ bosons}
Here, we show the derivation details of the simple case of $N=3$ particles for a physical Hamiltonian with only nearest-neighbor 1-body hoppings, for instance our $H_\text{1D}$ Hamiltonian introduced in the main text. We shall impose interactions with $n=1$ i.e. $t^\pm_\sigma = V^\pm_\sigma$, such that the excluded configuration space consists of the planes $x_1=x_2$, $x_2=x_3$ and $x_3=x_1$. 

The most general form for such a Hamiltonian (with lattice constant all set to unity) contains six real physical hoppings with amplitudes $t_1^\pm$, $t_2^\pm$ and $t_3^\pm$:
\begin{eqnarray}
H_\text{1D}^{N=3}(\bold k)&=& \sum_\pm t^\pm_1 e^{\pm ik_1}+t^\pm_2 e^{\pm ik_2}+t^\pm_3 e^{\pm ik_3}\notag\\
&=& \sum_\pm t^\pm_1 e^{\pm i(k_\text{CM}+2\tilde k_1+\tilde k_2)}+t^\pm_2 e^{\pm i(k_\text{CM}-\tilde k_1+\tilde k_2)}+t^\pm_3 e^{\pm i(k_\text{CM}-\tilde k_1-2\tilde k_2)}.
\label{poly0}
\end{eqnarray}
With generic values of the six hopping amplitudes, the GBZ and hence skin states will have to be found numerically, by diagonalizing Eq.~\ref{poly1} as a 2D lattice Hamiltonian with up to next-nearest-neighbor hoppings, where $k_\text{CM}$ enters as an external parameter. Combining all $k_\text{CM}$ sectors, we shall obtain the spectrum and full set of (PBC) skin eigenstates for $H_\text{1D}^{N=3}$.

What we have achieved is the solving of an interacting 3-body problem by mapping it to a single-body problem on a 2D non-Hermitian lattice. In general, this can always be done for an $N$-body CM-conserving interacting problem on a lattice, with the hopping asymmetries mapped onto the asymmetric hoppings of a $N-1$-dimensional lattice. Importantly, the physical requirement that the PBC skin state accumulations conserve CM  mandates that the accumulations be computed in a basis orthogonal to the CM translation direction, which in general do not coincide with the physical many-basis. As such, even single-particle nearest-neighbor hoppings can effectively assume the form of many-body hoppings ranging up to $N-1$ lattice sites, fundamentally altering the GBZ, as well as its analytic tractability. One salient emergent feature is a graph-like spectral structure in the complex plane~\cite{lee2019unraveling}, whose departure from the real line is physically rooted in the lack of convergence of certain non-Hermitian many-body processes.

\begin{figure}
\includegraphics[width=.9\linewidth]{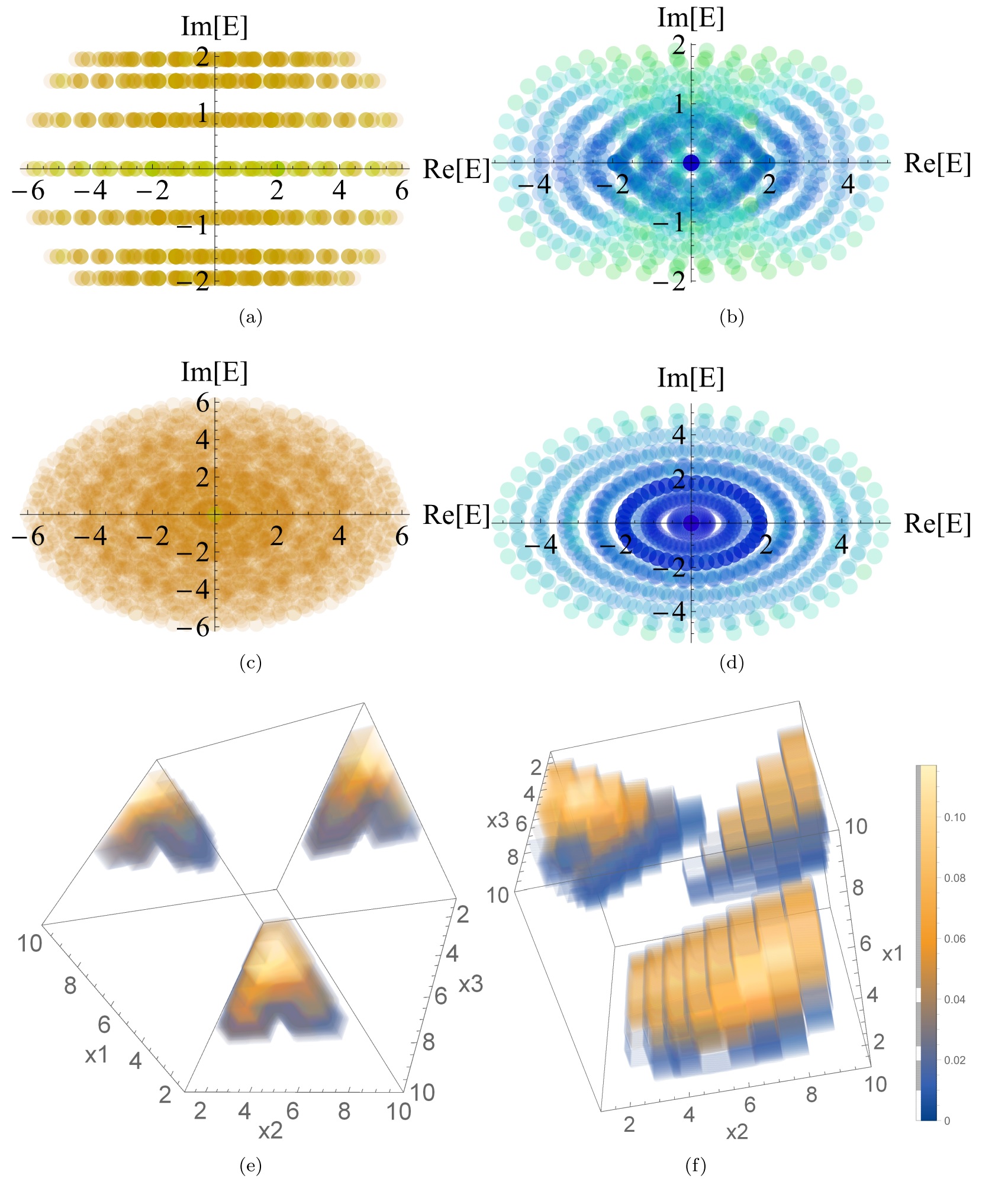}
\caption{(a,b) Complex spectra of $H_\text{1D}$ for $N=3$ bosons and parameter $t_1^\pm=t_2^\pm=1$, $t_3^-=0$, $t_3^+=2$, with interactions $V^\pm_\sigma$ turned off and on respectively. The PBC skin effect occurs even though only one of the particles experience asymmetric hoppings. (c,d) Complex spectra of $H_\text{1D}$ for $N=3$ bosons, for the case in Fig.~2c of the main text, also elaborated by Eq.~\ref{poly1}. (e,f) The profiles of arbitrarily selected eigenstates of the case in (d), at $E=-2.7-0.5i$ and $E=4.6-i$ respectively. Skin accumulation along the ``boundary'' planes $x_1=x_2$, $x_2=x_3$ and $x_3=x_1$ is always observed. $L=14$ was used throughout.}
\label{fig:N=3}
\end{figure}

\subsubsection{Analytic solution for some restricted cases of $H_\text{1D}^{N=3}$.}

\noindent We replace the LHS by the eigenenergy $E$, and consider the expression as a polynomial in $z_1=e^{i\tilde k_1}$ and $z_2=e^{i\tilde k_2}$ iteratively. WLOG, we start with $z_1$:
\begin{eqnarray}
E&=&  z_1^2 \left[t_1^+e^{i(k_\text{CM}+\tilde k_2)}\right]+\frac1{z_1^2} \left[t_1^-e^{-i(k_\text{CM}+\tilde k_2)}\right]+z_1\left[t_2^-e^{-i(k_\text{CM}+\tilde k_2)}+t_3^-e^{-i(k_\text{CM}-2\tilde k_2)}\right]+\frac1{z_1}\left[t_2^+e^{i(k_\text{CM}+\tilde k_2)}+t_3^+e^{i(k_\text{CM}-2\tilde k_2)}\right]\notag\\
\label{poly1}
\end{eqnarray}
The GBZ is generically determined by the smallest complex deformation $\kappa(\tilde p_1)$ such that $z_1\rightarrow e^{i\tilde p_1}e^{-\kappa(\tilde p_1)}$, $\tilde p_1\in \mathbb{R}$ gives rise to a double degeneracy in the solution for $E\rightarrow \bar E$ for each $\tilde p_1$. This shall be repeated for $z_2=e^{i\tilde k_2}$ too, after which the skin spectrum $\bar E\rightarrow \bar{\bar E}$ and eigenstates are obtained.

However, Eq.~\ref{poly1} is a degree four polynomial whose explicit solution is very complicated, even though it exists by the theorem of Abel-Ruffini. To make analytic headway, we shall consider a few cases where some of the hopping amplitudes vanish.
\begin{itemize}
\item \textbf{$t^\pm_1=0$ : }Having completely no hoppings for one of the particle species, chosen WLOG as species $1$, removes the quadratic powers of $z_1$ and $1/z_1$, leaving behind an effective 1D Hatano-Nelson model in the direction of $\tilde x_1$:
\begin{eqnarray}
E^{N=3}|_{t^\pm_1=0}&=&  z_1\left[t_2^-e^{-i(k_\text{CM}+\tilde k_2)}+t_3^-e^{-i(k_\text{CM}-2\tilde k_2)}\right]+\frac1{z_1}\left[t_2^+e^{i(k_\text{CM}+\tilde k_2)}+t_3^+e^{i(k_\text{CM}-2\tilde k_2)}\right]\notag\\
&=& z_1T^+_{23}+\frac1{z_1}T^-_{23}
\label{poly2}
\end{eqnarray}
where $T^\pm_{23}=t_2^\mp e^{\mp i(k_\text{CM}+\tilde k_2)}+t_3^\mp e^{\mp i(k_\text{CM}-2\tilde k_2)}$. With interaction-induced effective OBCs normal to $\tilde k_1$ (which takes the role of $p_\perp$) in the configuration space, the GBZ is given by $z_1\rightarrow e^{i\tilde p_1}\sqrt{\left|\frac{T^-_{23}}{T^+_{23}}\right|}=e^{i\tilde (p_1+\theta_{23})}\sqrt{\frac{T^-_{23}}{T^+_{23}}}$ where $\theta_{23}=\frac{\text{Arg}[T^+_{23}]-\text{Arg}[T^-_{23}]}{2}$, leading to
\begin{eqnarray}
E^{N=3}|_{t^\pm_1=0}\rightarrow \bar E^{N=3}|_{t^\pm_1=0}&=& e^{i(\tilde p_1+\theta_{23})}\sqrt{\frac{T^-_{23}}{T^+_{23}}}T^+_{23}+e^{-i(\tilde p_1+\theta_{23})}\sqrt{\frac{T^+_{23}}{T^-_{23}}}T^-_{23}\notag\\
&= & \sqrt{T^+_{23}T^-_{23}}\left(e^{i(\tilde p_1+\theta_{23})}+e^{-i(\tilde p_1+\theta_{23})}\right)\notag\\
&=&2\sqrt{T^+_{23}T^-_{23}}\cos(\tilde p_1+\theta_{23}),
\label{poly3}
\end{eqnarray}
which ranges between $-2\sqrt{T^+_{23}T^-_{23}}$ to $2\sqrt{T^+_{23}T^-_{23}}$, and is obviously immune to the NHSE in the $\tilde x_1$ direction. 

Constructing the GBZ in the second direction $\tilde k_2$ is however more subtle, since $\tilde k_2$ appears not just in $\sqrt{T^+_{23}T^-_{23}}$, but also in the $\theta_{23}$ angle in a way that is non-analytic at first sight. We would like to perform an analytic continuation on $\tilde k_2=-i\log z_2$ that brings the surrogate Hamiltonian into a form with balanced effective hoppings. Continuing from Eq.~\ref{poly3} with $T^+_{23}=e^{- i k_\text{CM}}(t^-_2/z_2+t^-_3z_2^2)$ and $T^-_{23}=e^{i k_\text{CM}}(t^+_2z_2+t^+_3/z_2^2)$,
\begin{eqnarray}
\bar E^{N=3}|_{t^\pm_1=0}&= &  \sqrt{T^+_{23}T^-_{23}}\left(e^{i(\tilde p_1+\theta_{23})}+e^{-i(\tilde p_1+\theta_{23})}\right)\notag\\
&=& \sqrt{t^-_2t^+_2+t^+_3t^-_3+t_3^-t_2^+z_2^3+t_3^+t_2^-/z_2^3}\left(\sqrt{\frac{T^+_{23}}{T^-_{23}}}\sqrt{\left|\frac{T^-_{23}}{T^+_{23}}\right|}e^{i\tilde p_1}+\sqrt{\frac{T^-_{23}}{T^+_{23}}}\sqrt{\left|\frac{T^+_{23}}{T^-_{23}}\right|}e^{-i\tilde p_1}\right).
\label{poly5}
\end{eqnarray}
To achieve balanced hoppings, we will at least need to balance the hoppings in the square root on the left, which resembles that of a a Hatano-Nelson model with lattice spacing of 3 units. We hence try the analytic continuation $z_2\rightarrow e^{i\tilde p_2}\sqrt[6]{\frac{t_2^-t_3^+}{t_2^+t_3^-}}$ with $\tilde p_2\in \mathbb{R}$, such that
\begin{eqnarray}
\sqrt{T^+_{23}T^-_{23}} &\rightarrow & \sqrt{t_2^+t_2^-+t_3^+t_3^-+2\sqrt{(t_2^+t_2^-)(t_3^+t_3^-)}\cos 3\tilde p_2 }
\label{poly6}
\end{eqnarray}
and
\begin{eqnarray}
\sqrt{\frac{T^+_{23}}{T^-_{23}}} &= &e^{-ik_\text{CM}}\frac1{z_2}\sqrt{\frac{t_2^-+t_3^-z_2^3}{t_2^++t_3^+/z_2^3}}\notag\\
 &\rightarrow &e^{-i(k_\text{CM}+\tilde p_2)}\sqrt[6]{\frac{t_3^-}{t_3^+}}\sqrt[3]{\frac{t_2^-}{t_2^+}}\sqrt{\frac{\sqrt{t_2^+t_2^-}+e^{3i\tilde p_2}\sqrt{t_3^+t_3^-}}{\sqrt{t_2^+t_2^-}+e^{-3i\tilde p_2}\sqrt{t_3^+t_3^-}}}\notag\\
 &= &\sqrt[6]{\frac{t_3^-}{t_3^+}}\sqrt[3]{\frac{t_2^-}{t_2^+}}\left[e^{-i(k_\text{CM}+\tilde p_2)}\frac{\sqrt{t_2^+t_2^-}+e^{3i\tilde p_2}\sqrt{t_3^+t_3^-}}{\sqrt{t_2^+t_2^-+t_3^+t_3^-+2\sqrt{(t_2^+t_2^-)(t_3^+t_3^-)}\cos 3\tilde p_2 }}\right]
\label{poly7}
\end{eqnarray}
with the quantity in the square parenthesis having unit modulus. It is interesting to note that the square root in the denominator of the last line of Eq.~\ref{poly7} is exactly $\sqrt{T^+_{23}T^-_{23}}$. Hence
\begin{eqnarray}
\bar{\bar E}^{N=3}|_{t^\pm_1=0}&= &  \frac{\sqrt{T^+_{23}T^-_{23}}}{\sqrt{T^+_{23}T^-_{23}}}\left(\left(\sqrt{t_2^+t_2^-}+e^{3i\tilde p_2}\sqrt{t_3^+t_3^-}\right)e^{-i(\tilde p_2+k_\text{CM}-\tilde p_1)}+\left(\sqrt{t_2^+t_2^-}+e^{-3i\tilde p_2}\sqrt{t_3^+t_3^-}\right)e^{i(\tilde p_2+k_\text{CM}-\tilde p_1)}\right)\notag\\
&=&2\sqrt{t_2^+t_2^-}\cos(\tilde p_2-\tilde p_1+k_\text{CM})+2\sqrt{t_3^+t_3^-}\cos(2\tilde p_2+\tilde p_1-k_\text{CM}).
\label{poly8}
\end{eqnarray}
Indeed, the energy of the skin states as given by Eq.~\ref{poly8} coincides with that of two independent Hatano-Nelson chains of hopping amplitudes $t_2^\pm$ and $t_3^\pm$. While this result can admittedly be obtained much more simply by treating $\tilde k_2-\tilde k_1+k_\text{CM}$ and $2\tilde k_2-\tilde k_1-k_\text{CM}$ as independent momenta from the outset (Eq.~\ref{poly2}), doing so will obscure the dissimilar skin accumulations between particles 1 and 2 vs 2 and 3. In our basis, the two orthogonal skin accumulation directions occur in the directions normal to $ \tilde x_1=x_1-x_2$ and $ x_2=x_2-x_3$, i.e. the relative displacements of particles $1,2$ and $2,3$ respectively. (The relatively displacement of particles $1,3$ is a simple linear combination $\tilde x_1$ and $\tilde x_2$.) Since particles $1$ cannot hop, unlike the other two particles, we will expect $\tilde \kappa_1$ and $\tilde \kappa_2$ to behave differently. 
Indeed, in the $\tilde x_2$ direction, the inverse skin depth is given by
\begin{equation}
\tilde\kappa_2=-\log|z_2|=\frac1{6}\log\left|\frac{t_2^+}{t_2^-}\right|-\frac1{6}\log\left|\frac{t_3^+}{t_3^-}\right|.
\label{kappa2}
\end{equation}
As expected, it is large either when $|t_2^+|>|t_3^+|$ and $|t_3^-|>|t_2^-|$ i.e. when the dominant hoppings of particles 2 and 3 are in opposing directions, resulting in a net accumulation between them. 

The $\tilde x_1$ direction accumulation, on the other hand, is characterized by the inverse skin depth (from Eqs.~\ref{poly2} and~\ref{poly7})
\begin{eqnarray}
\tilde \kappa_1=-\log|z_1|&=&\frac1{2}\log\left|\frac{T_{23}^+}{T_{23}^-}\right|\notag\\
&=&\frac1{2}\log\left|\frac{t_2^+z_2+t_3^+/z_2^2}{t_2^-/z_2+t_3^-z_2^2}\right|\notag\\
&=& \frac1{3}\log\left|\frac{t_2^+}{t_2^-}\right|+\frac1{6}\log\left|\frac{t_3^+}{t_3^-}\right|.
\end{eqnarray}
Here, since particle $1$ cannot hop, the skin accumulations solely arises from the left/right asymmetry of the hoppings of the other particles. Even though particle 3 is not directly involved in the interparticle displacement $\tilde x_1=x_1-x_2$, the fact that it enters the CM of all the particles, which needs to be conserved, causes its hoppings to also affect $\tilde \kappa_1$ to a smaller extent. 

\item \textbf{$t_1^-=t_2^+=t_3^+=0$ : } In this case, all the exponents of $z_1$ in Eq.~\ref{poly1} are positive, and the GBZ is degenerate, with all eigenvalues $\bar E$ collapsing onto zero via the so-called non-Bloch band collapse~\cite{PhysRevLett.124.066602}. This is because the separations between particle 1 and particles 2,3 can only decrease, not increase, as particle 1can only hop to the right while particles 2 and 3 can only hop to the left. With no counteracting hoppings, the skin accumulations degenerate such that all particles are right next to each other with $100\%$ probability.

\item \textbf{$t_1^-=t_2^-=t_3^-=0$ : } In this other nontrivial special case, all the (real) hoppings for the three species/particles are directed to the right. In other literature where OBCs are considered, all skin states will become degenerate, piling up on the rightmost site due to non-Bloch band collapse. But in our PBC scenario, having only rightwards hoppings do not lead to a trivial scenario, since the skin accumulations occur \emph{between} particles. The dispersion relation
\begin{eqnarray}
E^{N=3}|_{t^-_j=0}&=&  z_1^2 \left[t_1^+e^{i(k_\text{CM}+\tilde k_2)}\right]+\frac1{z_1}\left[t_2^+e^{i(k_\text{CM}+\tilde k_2)}+t_3^+e^{i(k_\text{CM}-2\tilde k_2)}\right]\notag\\
&=&z_1^2A+\frac{B}{z_1}
\label{poly4}
\end{eqnarray}
which is of the form of Eq.~\ref{EAB0} with $\alpha=2$, $\beta=1$ and $A,B$ given as above. As first derived in Ref.~\cite{lee2019unraveling} and representing a special case of Eq.~\ref{barE2}, we have
\begin{equation}
\bar E^{N=3}|_{t^-_j=0} \propto \sqrt[3]{AB^2}e^{\frac{2\pi i \nu}{3}},
\label{EAB9}
\end{equation}
which traces out a 3-spiked star in the complex plane, where $\nu$ labels the spikes. To derive $\bar{\bar E}^{N=3}|_{t^+_j=0}$, we attempt to balance
\begin{eqnarray}
AB^2&=&e^{3ik_\text{CM}}t_1^+z_2\left(t_2^+z_2+t_3^+/z_2^2\right)^2\notag\\
&=&e^{3ik_\text{CM}}t_1^+\left(t_2^+z_2^{3/2}+t_3^+/z_2^{3/2}\right)^2\notag\\
&\rightarrow& e^{3ik_\text{CM}}t_1^+\left(2\sqrt{t_2^+t_3^+}\cos\frac{3\tilde p_2}{2}\right)^2\notag\\
&\propto& e^{3ik_\text{CM}}t_1t_2t_3
\end{eqnarray}
via $z^3_2\rightarrow z^3_2\frac{t_3^+}{t_2^+}$, such that it takes the form of a Hatano-Nelson model with lattice constant being $3/2$ units. With this, we obtain the skin state eigenenergies which are proportional to
\begin{equation}
\bar{\bar E}^{N=3}|_{t^-_j=0}\propto e^{i(k_\text{CM}+2\pi\nu/3)}\sqrt[3]{t_1^+t_2^+t_3^+}.
\end{equation}
This result expresses how the eigenenergies scale on the whole with the geometric mean of the hoppings. The distribution of the individual eigenenergies, which are in principle indexed by $\tilde p_1$ and $\tilde p_2$, is numerically displayed in Fig.~2c of the main text.
As expected, all the nonzero hoppings enter symmetrically via $\sqrt[3]{t_1^+t_2^+t_3^+}$. In the next subsection, we shall see that this result generalized to $N$ particle species with only rightwards hoppings.

Since this eigenenergy is complex for $\nu\neq 0$, we expect states to undergo nontrivial attenuation/growth with time, with the fractional rate of change proportional to $\text{Im}\,\left|\bar{\bar E}^{N=3}|_{t^-_j=0}\right|\propto \frac{\sqrt{3}}{2}\sqrt[3]{t_1^+t_2^+t_3^+}$.

\end{itemize}

\subsection{Extension to certain scenarios with large number of particles $N$}

\begin{figure}
\subfloat[]{\includegraphics[width=.45\linewidth]{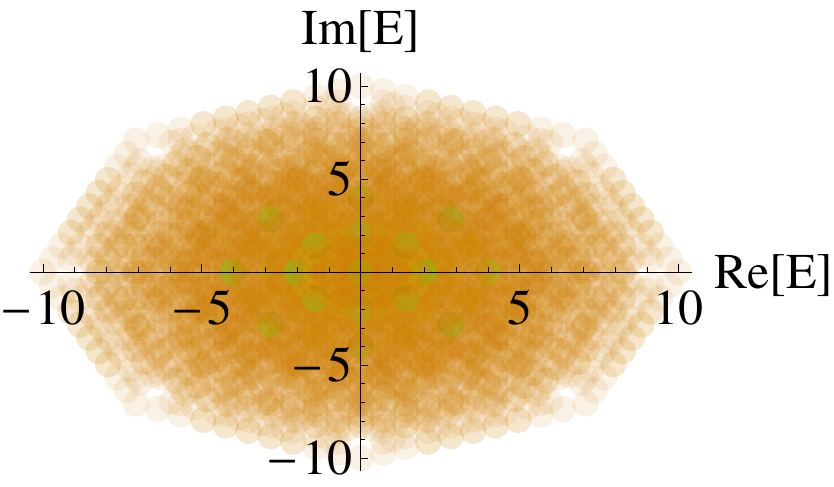}}
\subfloat[]{\includegraphics[width=.45\linewidth]{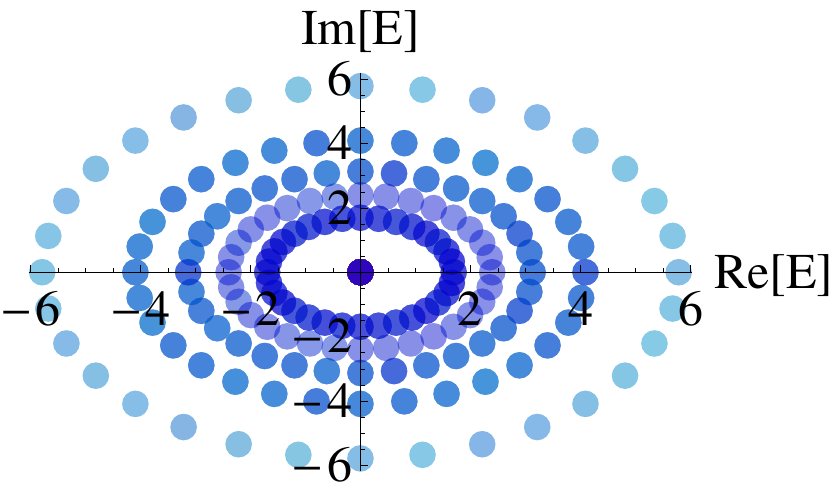}}\\
\subfloat[]{\includegraphics[width=.45\linewidth]{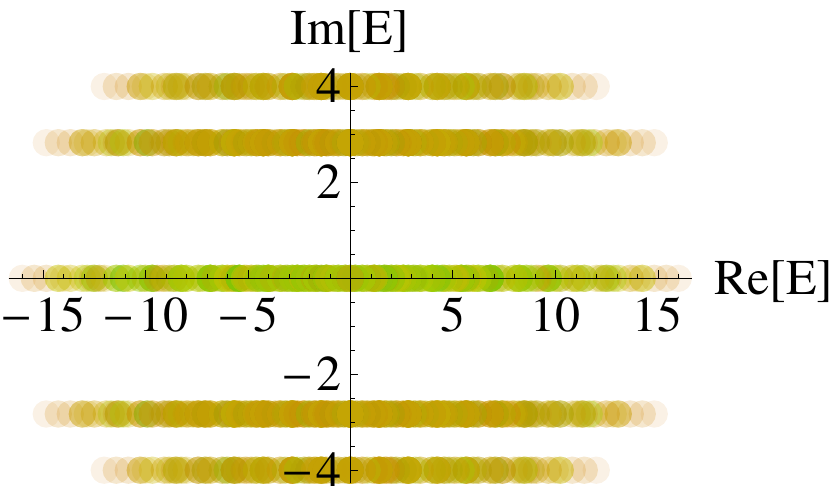}}
\subfloat[]{\includegraphics[width=.45\linewidth]{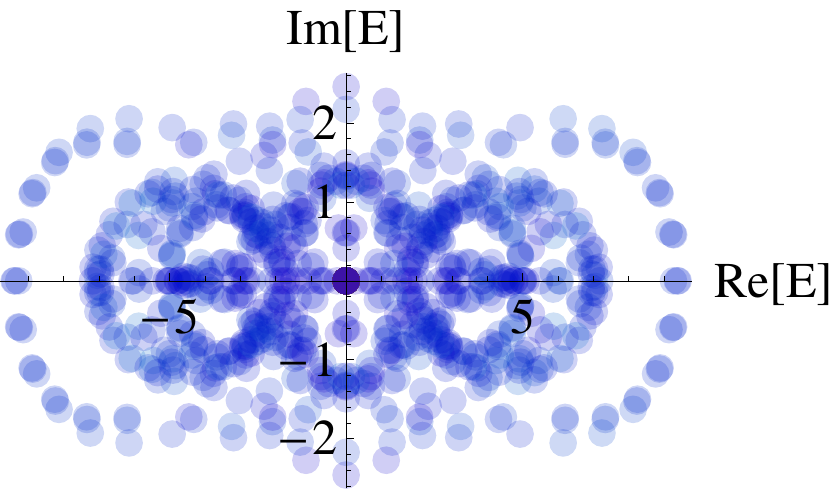}}\\
\subfloat[]{\includegraphics[width=.45\linewidth]{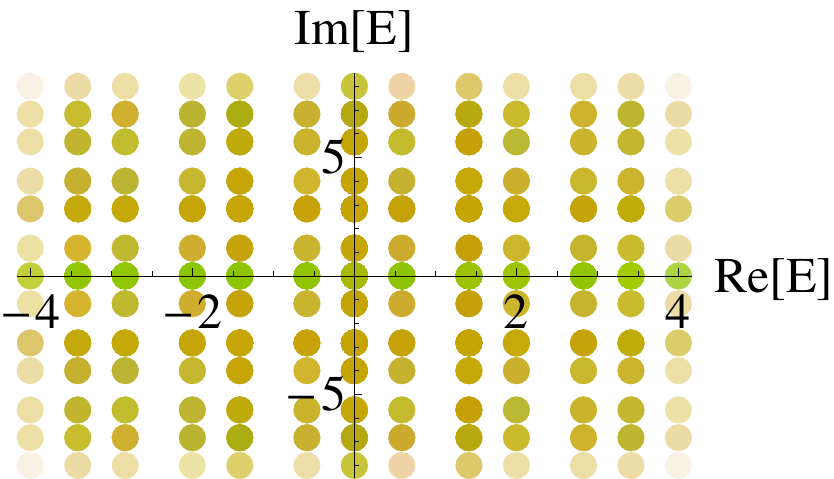}}
\subfloat[]{\includegraphics[width=.45\linewidth]{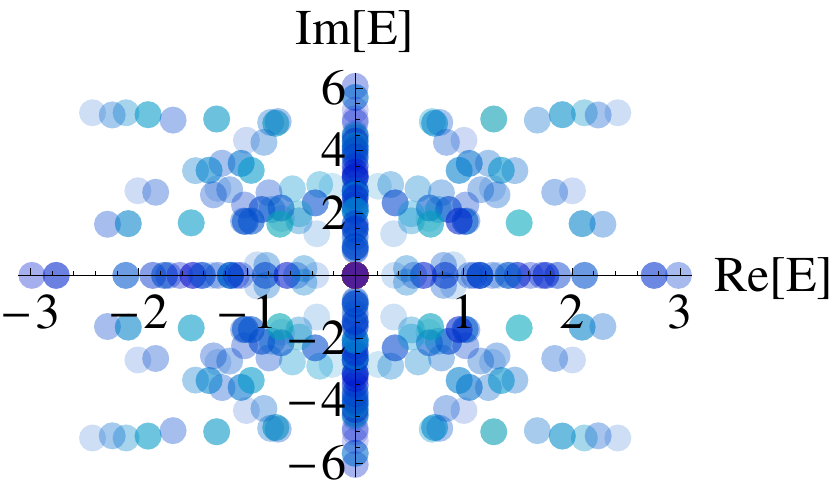}}
\caption{(a,b) Complex spectra of $H_\text{1D}$ for $N=4$ bosons with fully unidirectional hopping $t_j^+=j$, $t_j^-=0$ for $j=1,2,3,4$, with interactions $V^\pm_\sigma$ turned off and on respectively. This is qualitatively similar to the analogous unidirectional case in Fig.~\ref{fig:N=3} with $N=3$ particles.  (c,d) Complex spectra for the $N=4$ case in Fig.~2d of the main text, with interactions $V^\pm_\sigma$ turned off and on respectively. Indeed, the spectrum is modified dramatically by the interactions. (e-f) Complex spectra for another case with symmetric $t_1^\pm=t_3^\pm$, and almost symmetric $t_2^\mp=t_4^\pm=\pm 2$, with interactions $V^\pm_\sigma$ turned off and on respectively. A curious 4-fold symmetry emerges in the spectrum, and the non-interacting case has an almost uniform distribution of eigenenergies. $L=8$ was used throughout.
}
\label{fig:N=4}
\end{figure}

The thermodynamic (large $N$) limiting description of our system can be written down with the $M$ matrix results in the previous section, from Eq.~\ref{Minv} through Eq.~\ref{hop2}. Generically, it will involve $N-1$ iterated GBZ constructions for $\tilde k_1$ through $\tilde k_{N-1}$, and can only be performed numerically. However, here we show how the thermodynamic limit can be analytically accessed in the scenario with only unidirectional hoppings.

From the form of $M$ in Eq.~\ref{M} and the discussion that followed, our $N$-particle system with only nearest neighbor hoppings to the right can be described by
\begin{equation}
E^{N}|_{t^-_j=0}=e^{ik_\text{CM}}\sum_{j=1}^Nt_j^+\prod_{i=1}^{N-1}z_i^{M_{ij}}
\label{poly10}
\end{equation}
which for $N=3$ reduces to Eq.~\ref{poly4}: 
\begin{equation}
E^{N=3}|_{t^-_j=0}=e^{ik_\text{CM}}\sum_{j=1}^3t_j^+\prod_{i=1}^2z_i^{M_{ij}}=e^{ik_\text{CM}}\left(t_1^+z_1^2/z_2+t_2^+z_2/z_1+t_3^+/(z_1z_2^2)\right),
\end{equation}
where $z_i=e^{i\tilde k_i}$. Owing to the form of the matrix elements $M_{ij}=N\theta(i-j)-i$, at iteration $i$ where the GBZ of $z_{i}$ is being constructed, $M_{ij}=N-i$ for the first $i$ elements $M_{i1}$ to $M_{ii}$, and $M_{ij}=-i$ for the $i+1$-th to $N$-th elements $M_{i,i+1}$ to $M_{iN}$. 

Suppose we construct the GBZs in the order of $z_1,z_2,...,z_{N-1}$. For the first iteration, we have
\begin{eqnarray}
E^{N}|_{t^-_j=0}(z_1,z_2,...,z_{N-1})&=&e^{ik_\text{CM}}\left(z_1^{N-1}\left[t_1^+\prod_{i\neq 1}^{N-1}z_i^{M_{i1}}\right]+\frac1{z_1}\left[\sum_{j>1}^Nt_j^+\prod_{i\neq 1}^{N-1}z_i^{M_{ij}}\right]\right)\notag\\
&\rightarrow & \bar{E}^{N}|_{t^-_j=0}(z_2,...,z_{N-1})\notag\\
&\propto& e^{ik_\text{CM}}\left[t_1^+\prod_{i\neq 1}^{N-1}z_i^{M_{i1}}\right]^{\frac1{N}}\left[\sum_{j>1}^Nt_j^+\prod_{i\neq 1}^{N-1}z_i^{M_{ij}}\right]^{\frac{N-1}{N}}\notag\\
&= & e^{ik_\text{CM}}\left[t_1^+\prod_{i\neq 1}^{N-1}z_i^{N-i}\right]^{\frac1{N}}\left[\sum_{j>1}^Nt_j^+\prod_{i\neq 1}^{N-1}z_i^{N\theta(i-j)-i}\right]^{\frac{N-1}{N}}\notag\\
&= & \left(e^{ik_\text{CM}}t_1^+\right)^\frac1{N}\left[e^{ik_\text{CM}}\sum_{j>1}^Nt_j^+\prod_{i> 1}^{N-1}z_i^{\frac{N}{N-1}(N\theta(i-j)-i)}\right]^{\frac{N-1}{N}}\notag\\
&\rightarrow & \left(e^{ik_\text{CM}}t_1^+\right)^\frac1{N}\left[E^{N-1}|_{t_j^-=0}(z_2,...,z_{N-1})\right]^{\frac{N-1}{N}}
\label{poly10}
\end{eqnarray}
The first $\rightarrow$ comes about from the GBZ construction of $z_1$, which results in a modified spectrum $\bar{E}^{N}|_{t^-_j=0}(z_2,...,z_{N-1})$ computed via Eq.~\ref{barE2}. The second $\rightarrow$ involves the replacement $z_i\rightarrow z_i^{(N-1)/N}$ for all $i>1$, which amounts to a rescaling of the unit cell by the factor $(N-1)/N$. We have thus shown that, after one iteration of GBZ construction, the $N$-particle spectrum is proportional to the $(N-1)/N$-th power of the $N-1$-particle spectrum with particle 1 omitted. Applying this procedure repeatedly, we finally arrive at
\begin{eqnarray}
\bm\bar{E}^{N}|_{t^-_j=0}\propto e^{ik_\text{CM}}\sqrt[N]{t_1^+t_2^+...t_N^+}e^{2\pi i \nu/N}.
\label{poly11}
\end{eqnarray}
This result shows that the spectrum scales overall with the geometric mean of all the hoppings, consistent with symmetry. This finding is nonetheless nontrivial, since there exist various other symmetric polynomials that could have been involved. As with the 3-body case, the detailed distribution of individual eigenenergies is much more complicated and can in most cases only be obtained numerically.

Note that Eq.~\ref{poly11} also implies that, if there are only a finite number of species $s$ with respective particle number proportions $f_1,f_2,...,f_s$, each experiencing only hoppings $t_i$ in the same direction, the skin spectrum will scale with
\begin{eqnarray}
\bm\bar{E}^{N}|_{t^-_j=0}\propto e^{ik_\text{CM}}\prod_{j=1}^st_j^{f_j}e^{2\pi i \nu/N},
\label{poly12}
\end{eqnarray}
which implies a spectral distribution independent of the total number of particles.

\subsection{Cases with partial ``boundaries''}

In the main text, we have explicitly analyzed only cases where interactions or particle statistics fully render certain configurations inaccessible. But realistic interactions cannot be perfectly tuned. Furthermore, partial suppression of hoppings also enrich the set of possible dynamics.

\begin{figure}
\subfloat[]{\includegraphics[width=.33\linewidth]{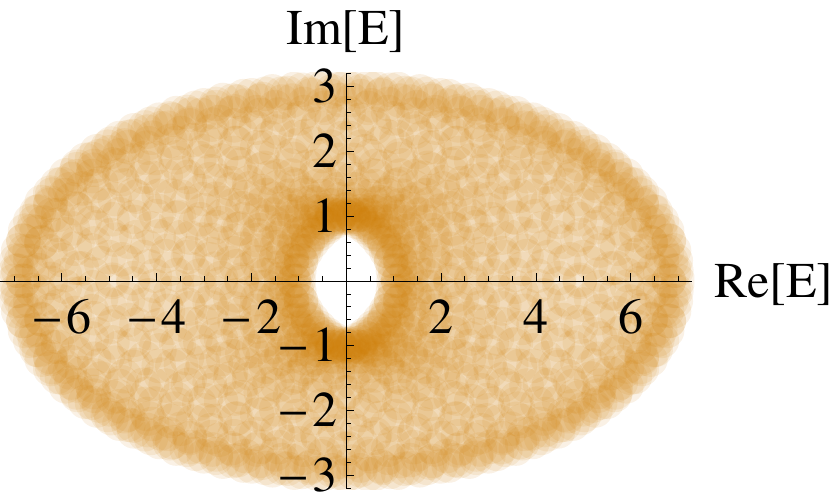}}
\subfloat[]{\includegraphics[width=.33\linewidth]{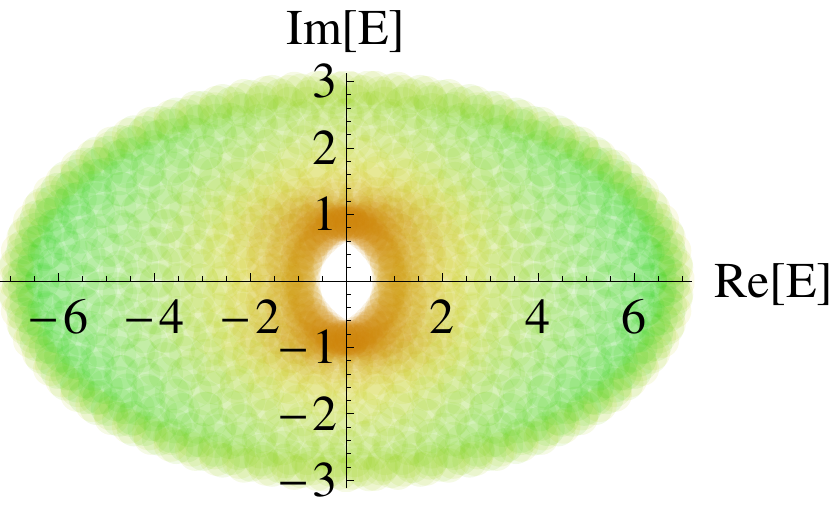}}
\subfloat[]{\includegraphics[width=.33\linewidth]{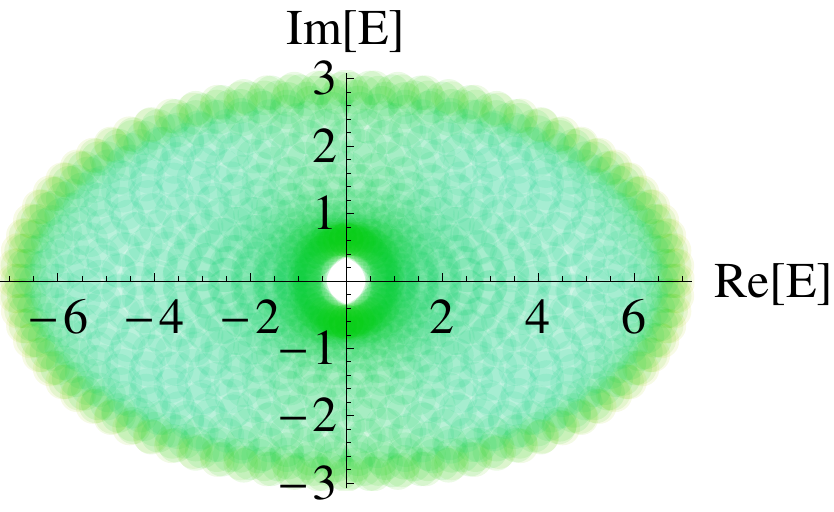}}\\
\subfloat[]{\includegraphics[width=.33\linewidth]{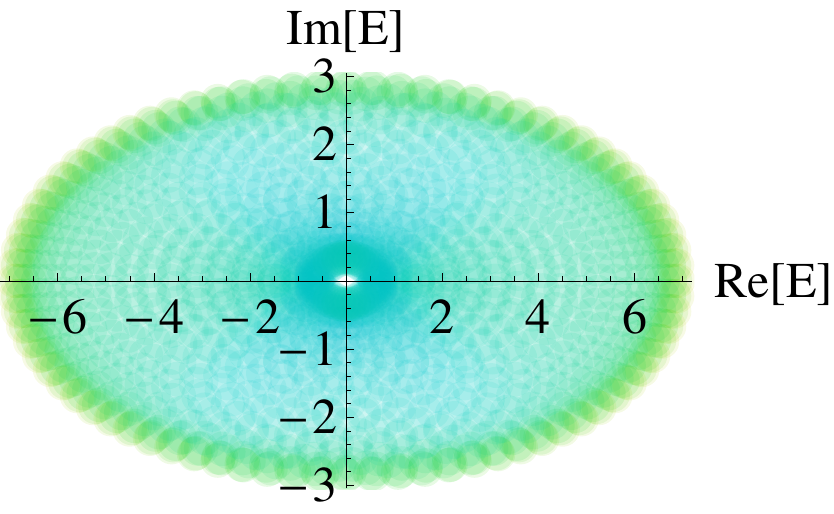}}
\subfloat[]{\includegraphics[width=.33\linewidth]{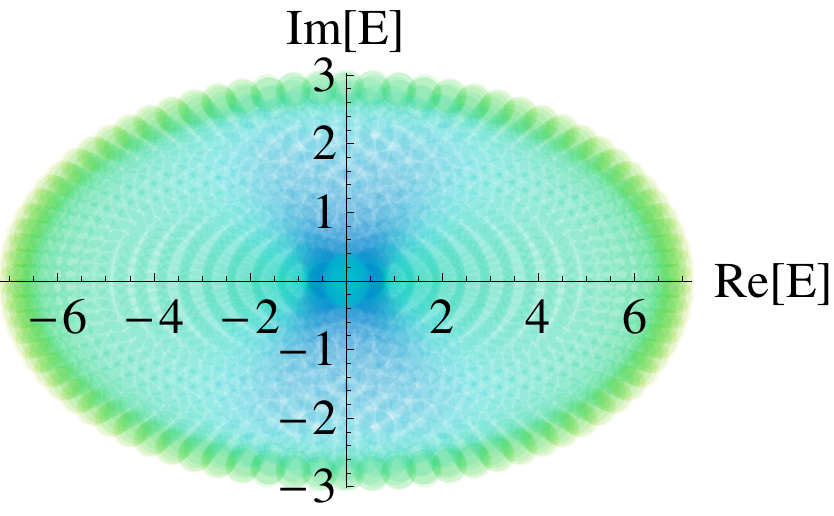}}
\subfloat[]{\includegraphics[width=.33\linewidth]{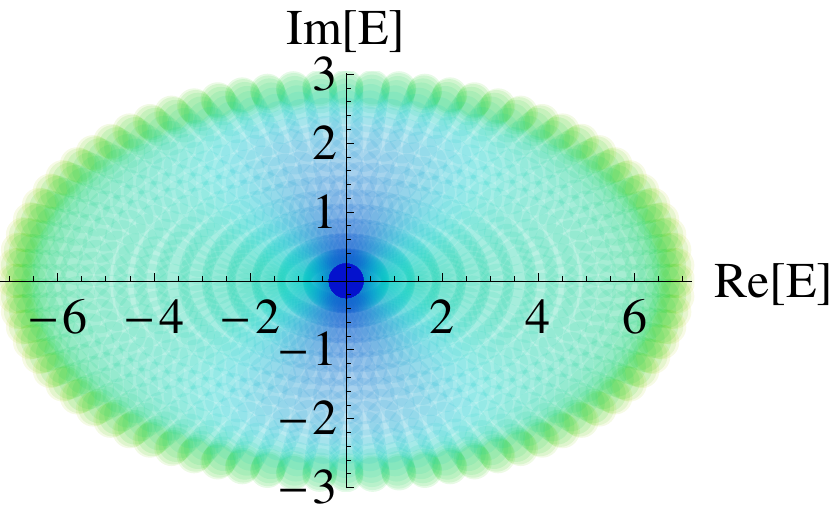}}
\caption{Complex spectra for $H_\text{1D}$ with $N=2$ particles for hopping strengths $t_a^-=t_b^-=1$, $t_a^+=2$, $t_b^+=3$. From (a) to (f), we $n=t^{\pm}_\sigma/V^\pm_\sigma =1/(1-\Delta)$ with $\Delta$ ranging across $\Delta=1$ (no interactions), $10^{-1}$, $10^{-2}$, $10^{-3}$, $10^{-4}$ to $0$ (full interaction that prevents double occupancy). The spectrum resembles the $\Delta=1$ case even for a relatively small $\Delta$ of $0.01$ to $0.1$.
}
\label{fig:N=2_interpol}
\end{figure}

In general, the incomplete attenuation of hoppings can be systematically studied by interpolating between full PBCs and OBCs of the configuration space. In the non-Hermitian setting, this can be shown to correspond to the spectral flow generated by an imaginary flux~\cite{lee2019anatomy}, at least up to a certain extent. The key takeaway is that the OBC spectrum is exponentially sensitive, such that even a small perturbation is likely to make the spectrum qualitative resemble the PBC limit. Shown in Fig.~\ref{fig:N=2_interpol} is the spectrum of $H_\text{1D}$ with $N=2$ particles [Eq.~1 of main text] with asymmetric $t^-_a=1,t^+_a=2,t^-_b=1,t^+_b=3$, such that the interaction $V^\pm_\sigma$ successively approaches exponentially close to $t^\pm_\sigma$, such that the net hopping towards an occupied site $\propto\Delta$ exponentially approaches zero. 

As such, the hoppings need to be attenuated to at least an order of magnitude smaller than their original amplitude before they can behavior like ``boundary'' hoppings. Note, however, that our discussion is based on a scenario where high densities suppress hoppings, which is the case for our $H_\text{1D}$ at sufficiently low densities, even if $n$ deviates from an integer value. At higher densities, $H_\text{1D}$ should be replaced by a more phenomenologically realistic nonlinear model that always suppresses hoppings towards densely populated sites.

\section{Chiral topological states from PBC 2-body hopping model}

We start from 
\begin{eqnarray}
H_\text{1D}^\text{topo}&=&t\sum_{\Delta x_1,\Delta x_2=\pm 1}\mu^\dagger_{x_1+\Delta x_1}\nu^\dagger_{x_2+\Delta x_2}\mu_{x_1}\nu_{x_2}e^{i\lambda\phi|\Delta x_1+\Delta x_2|}+t'\,\sum_{x}\lambda (\mu^\dagger_{x+2}\mu_{x}-\nu^\dagger_{x+2}\nu_{x})+\text{h.c.}
\end{eqnarray}
of the main text [see Fig.~3 therein], which can be interpreted as a 2-fermion hopping model on a 1D zigzag lattice with PBCs. The even/odd sites of the chain correspond to the top/bottom levels of the zigzag chain, which are labeled $\lambda=(-1)^x$. $\mu,\nu$ refers to the 2nd-quantized operators of the two fermions. In anticipation of obtaining a Checkerboard geometry for the configuration space, we design the hoppings such that the fermions are on the same level of the zigzag chain at any one time i.e. $x_1+x_2 \equiv 0\, (\text{mod } 2)$: they should not simultaneously occupy an odd and an even site. 

There are 4 possible 2-body hopping processes of amplitudes $t$, with each fermion either hopping left or right. When the two fermions hop in the same direction, a flux from a vertical time-reversal breaking field gives rise to a phase factor of $e^{\pm 2i\phi}$, the sign depending on whether the two fermions were initially on the upper or lower level. When the fermions hop in opposite directions, no phase is incurred (As we see later, the topology remains stable even if small asymmetric phases are accrued in these cases, but having these phases make the system less realistic). 

Additionally, there are also minimal 1-body hoppings $\pm t'$ across two sites (1-site single-body hoppings will result in $x_1+x_2 \equiv 1\, (\text{mod } 2)$). To open up the bulk gap, we stipulate that the hopping is $+t$ when fermion $\mu$($\nu$) is on the upper(lower) level, and $-t$ vice versa.

\begin{figure}
\includegraphics[width=\linewidth]{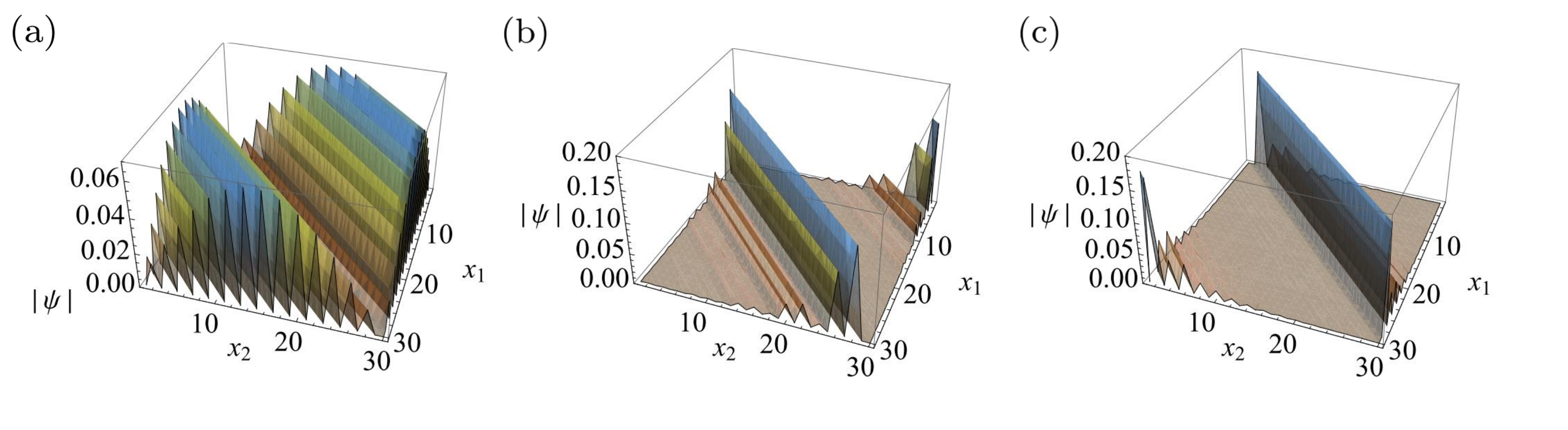}
\caption{Profiles of representative eigenstates in the 2-body configuration space, obtained by diagonalizing $H_\text{1D}^\text{topo}$ (and not $H_\text{config}^\text{topo}$ directly). (a) Somewhat delocalized $E=2$ at the edge of the bulk gap, (b) in-gap eigenstate at $E=1.7$, which is localized against the $x_1=x_2$ ``boundary'', (c) even more strongly localized eigenstate at $E=1.2$. }
\label{fig_CB}
\end{figure}

Putting this system into its 2-body configuration space, and defining $k_\perp = k_1-k_2$, $k_\text{CM}=k_1+k_2$, the system takes the form of a Checkerboard lattice [Fig.~3a of the main text] with $t'$ taking alternate signs in adjacent ``Checkers'', and flux of alternating signs along the $CM$ direction. Since this system is periodic in both $x_1$ and $x_2$, the Hamiltonian can be expressed in momentum space as 
\begin{eqnarray}
H_\text{config}^\text{topo}(\bold k)&=&2t(\cos k_\perp+\cos k_\text{CM}\cos 2\phi)\sigma_x-2t\sin 2\phi\cos k_\text{CM}\sigma_y-4t'\sin k_\perp \sin k_\text{CM}\sigma_z.
\label{Hconfig}
\end{eqnarray}
As plotted in Figs.~3b,c of the main text, with ``open boundaries'' perpendicular to $k_\perp$ arising from Pauli exclusion that prohibits occupancy of $x_1=x_2$ states, 2 sets of localized mid-gap states appear inside the bulk gap. These chiral modes traverse the bulk gap twice in one period of $k_\text{CM}$, signifying a Chern number of $2$ for the lower occupied band, which is also corroborated by the integral of the Berry curvature distribution. The in-gap chiral modes are well-localized with $\text{IPR}<0.1L$, and are displayed explicitly in Fig.~\ref{fig_CB}. 

The finite dispersions of the chiral states with respect to $k_\text{CM}$ implies nonzero uni-directional group velocities of wavepackets near $x_1=x_2$. In the physical zigzag chain, they correspond to states with the two fermions clustered close to each other. While the chiral transport of ordinary Chern modes can be understood through a flux pumping argument across physical edges, in our case, there are no edges, only dynamical ``boundaries'' set by one impenetrable fermion on another. Although the hoppings in $H_\text{1D}^\text{topo}$ appear symmetric, they are not when one particle is near the other, since certain hoppings will be forbidden by fermionic statistics. As one fermion cannot be on both sides of another at the same time, this hopping suppression becomes  intrinsically asymmetric, leading to the emergence of chiral propagation.

\end{document}